\documentclass[final,5p,times,twocolumn,authoryear]{elsarticle}

\usepackage{amssymb}
\usepackage[T1]{fontenc}
\usepackage[utf8]{inputenc}
\usepackage{amsmath}
\usepackage{graphicx}
\usepackage{subfigure}
\usepackage{latexsym}
\usepackage[normalem]{ulem}
\usepackage{rotating}
\usepackage{multirow} 

\usepackage[
  citecolor=blue,
  colorlinks,
  linkcolor=blue,
  urlcolor=blue,
]{hyperref}
\usepackage[dvipsnames]{xcolor}

\begin{document}

\begin{frontmatter}

\author[iitk]{Saikat Sur\corref{cor1}}
\ead{saikatsu@iitk.ac.in}
\author[iitk]{Anupam Ghosh}
\ead{anupamgh@iitk.ac.in}

\cortext[cor1]{Corresponding author}

\address[iitk]{Department of Physics, Indian Institute of Technology Kanpur, 
Uttar Pradesh 208016, India}

\title
{Quantum counterpart of Measure synchronization: A study on a pair of Harper systems}

\author{}

\address{}

\begin{abstract} 
Measure synchronization is a well-known phenomenon in coupled classical Hamiltonian systems over last two decades. In this paper, synchronization for coupled Harper system is investigated in both classical and quantum contexts. The concept of measure synchronization involves with the phase space and it seems that the measure synchronization is restricted in classical limit. But, on the contrary, here, we have extended the aforesaid synchronization in quantum domain. In quantum context, the coupling occurs between two many body systems via a time and site dependent potential. The coupling leads to the generation of entanglement between the quantum systems. We have used a technique, which is already accepted in the classical domain, in both the contexts to establish a connection between classical and quantum scenarios. Interestingly, results corresponding to both the cases lead to some common features.  
\begin{description}
\item[PACS numbers]{05.45.Xt}, {05.45.{-}a}, {03.65.Ud}, {03.67.Bg}, {03.67.Hk}, {75.10.Pq}
\end{description}
\end{abstract}


\end{frontmatter}


\section{Introduction}
\label{sec:intro}
Synchronization was first observed in the seventeenth century by
Huygens~\citep{huygens73}, while working with the coupled pendulum clocks. Later, in
the twentieth century, it was also reported that this phenomenon is observed in a triode
generator, in two organ pipes, etc.~\citep{prk2001}. In addition, the lighting of
fireflies~\citep{agu2011} is an example that supports the existence of this
phenomenon in nature. Also, in the extended ecological systems~\citep{bla1999} and in
the metabolic process~\citep{shliom2014} we observe this phenomenon. However, much later, synchronization for the coupled chaotic systems was observed and chaotic synchronization becomes popular after $1990$~\citep{pc1990}.
Chaotic synchronization was challenging because of the sensitive dependence on the
initial conditions. However, further progress in this field revealed the existence
of the different kinds of synchronization, viz., phase synchronization, generalized
synchronization, lag synchronization, etc.~\citep{bocc02}. In delayed systems also,
this phenomenon is observed~\citep{pc15}. Though there is a vast literature for
synchronization in coupled dissipative~\citep{pc15,ghosh18} and coupled delay chaotic systems~\citep{bocc02}, few articles are
reported regarding the synchronization of Hamiltonian chaos. In the coupled
Hamiltonian systems, the aforesaid synchronizations are not observed. Also, since the
Hamiltonian systems follow the Liouville's theorem, there is no existence of
attractor here unlike the dissipative systems. Therefore, one
would get a different type of synchronization: `measure synchronization' (MS), in coupled Hamiltonian systems, reported
in $1999$~\citep{hamp99}.
In case of MS, the participating subsystems share the same
identical measure in the projected phase space~\citep{hamp99}. For two coupled Hamiltonian
systems, if both the subsystems share the same area in the projected phase
portraits, then the subsystems are in MS state, otherwise, not. Further, it is also reported that when there are more than two subsystems---say three subsystems---it may be observed that the first subsystem is in MS state with the third one, but not with the second one---this kind of synchronization is called the partial measure synchronization~\citep{vincent05}. \textcolor{black}{MS is studied in various branches of Physics, viz., in bosonic Josephson junction~\citep{tian13}, in delayed coupled Hamiltonian systems~\citep{saxena13}, in ultra-cold atomic clouds~\citep{qiu15}, in optomechanical systems~\citep{bemani17}, etc.}
Now, since synchronization in the coupled Hamiltonian systems is observed, one may
extend the concept of the MS in the quantum systems. But, the
first problem we face regarding this issue, is the fundamental idea behind the MS involves with the concept of the region covered
by the phase space trajectory; there is no concept of phase space in quantum mechanics. However, recently, one article~\citep{qiu2014} reported the quantum many body measure
synchronization, and they have shown that both the coupling subsystems have identical measure in the three dimensional angular momentum space. MS has also been studied in quantum many body systems \citep{qiu10,tian} by extending the definition classical MS to quantum systems. Recently a study has been done on synchronization and quantum entanglement generation \citep{roulet}. 
Quantum entanglement is a unique property of quantum mechanics. There is no analogue of entanglement in classical mechanics. Due to the linear superposition principle and tensorial structure in quantum mechanics, certain quantum state of two or more than two subsystems can not be written as a product of the states of individual subsystems. Hence, the subsystems can have quantum correlations even they are spatially separated. This results in certain phenomena which are exclusively quantum in nature \citep{epr}. Quantum entanglement has been studied in various contexts and recognized as an essential resource for quantum computation and information in last few decades. 
Many body quantum systems specially quantum spins are of great interest in quantum communication theory  as they could be used as quantum wires to join quantum devices. So quantum entanglement has also been extensively studied in the field of communication in quantum spin systems, viz., state transfer, entanglement transfer in quantum spin chains \citep{bose,vs04,christandl}, etc. To establish a connection between the MS and the entanglement generation in many-body quantum systems, one needs to have a quantum system which behaves as a classical system  showing measure synchronization in certain limit. Previously some studies show the connection between entanglement generation in higher dimensional quantum systems and chaos in their classical  map \citep{arul01}.
\textcolor{black}{Here, we concentrate on a one dimensional periodically kicked Harper model with $N$ qubits, i.e., $2^N$ dimensional Hilbert space. This model reduces to a classical Harper map for large $N$ limit. This is an approximate model for electrons in a crystal lattice subjected to a perpendicular magnetic field~\citep{harper}. The dynamics of Bloch electrons moving in periodic potential in presence of a uniform magnetic field and time varying electric field can be shown to be governed by Harper dynamics under the tight binding approximation~\citep{iomin1,iomin2}. The model has a rich spectral structure, and also relevant in the context of metal-insulator transition or transition from localized to extended states~\citep{basu_1991,artuso94, artuso2}. In addition, the quantum kicked Harper model has been investigated for the entanglement distribution and dynamics~\citep{lakshminarayan03}.} 
Since there is no closed form solution for the quantum model due to the space dependent on site potential we restrict ourselves in one particle subspace. However, it is possible to write time dependent wave-function for one particle states in analytical form as discussed in~\citep{sur18_2}. Coupling such two $N$ qubit systems via another time and space dependent potential can result in interesting consequences. This leads to decoherence in individual systems and both the systems get correlated. Due to the absence of phase space in quantum mechanics, their coupling dynamics can be investigated by studying their quantum correlation measures as well as intra qubit correlations from the two systems. 
In this paper, we study the coupled Harper systems and try to observe the measure
synchronization here. Since the concept of MS is purely classical, we try to study the coupled Harper systems in the classical limit. Further, we show that one can extend the idea of MS in the domain of coupled quantum systems. Thus, this manuscript aims to connect two completely different branches of physics, synchronization in chaotic systems and quantum many-body physics and intrigues some new avenues for further research. The manuscript is prepared in the following order: In section III, we explain the MS explicitly in the classical scenario, then we return to the quantum picture and study the dynamics in detailed in section IV, and try make an analogy. Finally, a short discussion on local coupling in quantum scenario has been added.   
\section{Preliminaries}
We consider the following integrable Hamiltonian~\citep{lakshminarayan03} of a one dimensional  $N$ body quantum system  of fermions hopping on a chain with an inhomogeneous site potential. The Hamiltonian is given by:
\begin{equation}
\label{eq:quan_hamil}
{H} = \frac{1}{2} \sum^N_{j= 1} {c}^\dagger_j {c}_{j+1} + \frac{g}{2} \sum^N_{j= 1} {d}^\dagger_j {d}_{j+1} + \rm{H.C.},
\end{equation}
where $c_j^\dag$ are the creation operators at site $j$, $g$ is the potential strength parameter. The operators ${d_k}$ are the Fourier transformation of the Fermion annihilation operators and are given as follows:
\begin{equation}
{d}_k = \frac{1}{\sqrt{N}} \sum^N_{j= 1} e^{\frac{2\pi i jk}{N} } {c}_{j}.
\end{equation}
Substituting it in the actual Hamiltonian we get:
\begin{equation}
{H} = \sum^N_{j= 1} \big[\frac{1}{2} ( {c}^\dagger_j {c}_{j+1} + {c}_j {c}^\dagger_{j+1}) + g \cos(\frac{2 \pi j}{N}){c}^\dagger_j{c}_j\big].
\end{equation}
Further, we denote the sum of the operators in the following way:
\begin{eqnarray}
\label{eq:uv}
&{V} =  \sum^N_{j= 1} {c}^\dagger_{j+1} {c}_{j}, \nonumber\\
&{U} =  \sum^N_{k= 1} {d}^\dagger_{k} {d}_{k+1}  = \sum^N_{j=1} e^{\frac{2 \pi ij}{N}} {c}^\dagger_j {c}_j,
\end{eqnarray}
and, it can be shown that the operators ${V}$ and ${U}$ are the discrete versions of standard quantum position and momentum translation operators: $\exp(-i{p}a/\hbar)$ and $\exp(-i{x}b/\hbar)$ respectively. Here, $a$ and $b$ are respectively the smallest position and the momentum units. A detailed explanation is given in~\citep{lakshminarayan03}. However, in terms of this  new notation the Hamiltonian can be written as:
\begin{eqnarray}
\label{eq:inte_hamil}
{H} = \frac{1}{2} \big[ {V} +{V}^{\dagger}  \big]  + \frac{g}{2} \big[ {U} +{U}^{\dagger}  \big] \nonumber\\
=  \frac{1}{2} \big[ {V} +{V}^{\dagger}  \big]  + \frac{g}{2} \sum^N_{j= 1}\cos(\frac{2 \pi  j}{N}){c}^\dagger_j{c}_j ,
\end{eqnarray}
Now, we consider the non-integrable Hamiltonian, i.e., the second term on the right hand side of Eq.~\ref{eq:inte_hamil}---which can also be thought of as a time dependent and site dependent potential term. The non-integrable Hamiltonian is given as: 
\begin{eqnarray}
\label{eq:inte_hamil_t}
&{H}(t)=\frac{1}{2} \big[ {V} +{V}^{\dagger}  \big]  + \frac{g}{2} \big[ {U} +{U}^{\dagger}\big]   \sum^{\infty}_{n =-\infty} \delta(\frac{2\pi t }{\tau} -n) \nonumber\\
&=\sum^N_{j= 1} \big[\frac{1}{2} ( {c}^\dagger_j {c}_{j+1} + {c}_j {c}^\dagger_{j+1}) + g \cos(\frac{2 \pi j}{N}){c}^\dagger_j{c}_j \nonumber\\&\sum^{\infty}_{n =-\infty}  \delta(\frac{2\pi t }{\tau} -n)\big]. \nonumber\\
\label{fermion_Hamiltoian}
\end{eqnarray}
A train of impulses is provided at intervals of time $\tau/(2 \pi)$, where $\tau$ is the kicking interval parameter. As $\tau \rightarrow 0$, we recover the integrable Harper equations. The first term represents the kinetic energy of the fermion or hopping term, and the second term is the kicked potential energy operator.   The effect of the potential is through a train of kicking pulses with an interval $\tau$, a tunable parameter, to go continuously from completely integrable to completely non-integrable regimes. For  $\tau \rightarrow 0$ the dynamics of the Harper Hamiltonian is integrable, and for large values of $\tau$ the dynamics is completely chaotic (see Fig.~$4$ in the work of Lakshminarayan and Subrahmanyam~\citep{lakshminarayan03}, and for further details~\citep{lima91}). The potential strength parameter $g$ and the kicking interval parameter $\tau$ can be varied  independently to change the dynamics qualitatively.
\textcolor{black}{The Hilbert space for a single site is two-dimensional, either occupied or unoccupied, and thus it can be mapped to the spin language. We convert the Hamiltonian in Eq.~\ref{fermion_Hamiltoian} from fermion operator to spin operator formalism via Jordan-Wigner transformation~\citep{lieb61}, where the fermion occupation is mapped to the down spin occupation in the spin states. The Hamiltonian can be written in terms of spin operator language as the following,}
\textcolor{black}{\begin{eqnarray}
	\label{eq:hamil_a}
	&{H}(t)=\sum^N_{j= 1} \big[\frac{1}{4} [({\sigma}^x_{j}{\sigma}^x_{j+1}+{\sigma}^y_{j}{\sigma}^y_{j+1})\nonumber\\&+g \cos(\frac{2\pi j}{N}) \frac{{\sigma}^z_{j} +1}{2} \sum_{n=- \infty}^{\infty} \delta(\frac{2\pi t}{\tau}-n)].
	\end{eqnarray}}
\textcolor{black}{The first term turns out to be XY term, and the second term becomes a transverse field that is inhomogeneous in space.  This incorporates an interaction of down spins on neighbouring sites and there is no many-body interaction here.}
Now, we discuss the coupling scheme of two identical Harper systems, denoted by $A$ and $B$ respectively, each with $N$ particles, The  time dependent coupling Hamiltonian is proposed by taking the product of potential terms from individual Hamiltonians. This coupling scheme is very much similar to~\citep{miller99}. The time dependent Hamiltonian for the coupling term are given by the following,

\begin{eqnarray}
\label{eq:hamil_ab}
&{H}^{AB}(t) = \frac{\varepsilon}{2}\big[ {U}^{A} +{U^{A}}^{\dagger}\big] \big[ {U}^{B} +{U^{B}}^{\dagger}\big]    \sum^{\infty}_{n =-\infty} \delta(\frac{2\pi t }{\tau} -n),\nonumber\\
\end{eqnarray}
where $\varepsilon$ is the coupling strength parameter. The coupling strength parameter $\varepsilon$ should be of same order of  the potential strength parameter $g$. We set $g$ unity throughout the paper and varied $\varepsilon$ to see the measure Synchronization.
%

\textcolor{black}{Using the definition for the operator $U$  given in Eq. \ref{eq:uv}, we can write the coupling Hamiltonian in terms of the spin operators,}
\begin{eqnarray}
\label{eq:hamil_ab2}
{H}^{AB}(t) =  \frac{\varepsilon}{2}\sum^N_{j_A,j_B=1} \cos(\frac{2 \pi j_A}{N}) \cos(\frac{2 \pi j_B}{N})\sigma^z_{j_A}\sigma^z_{j_B}\nonumber\\ \sum^{\infty}_{n =-\infty} \delta(\frac{2\pi t }{\tau} -n).\nonumber\\
\end{eqnarray}
Rest of the terms vanish because of the transverse potential being symmetric in site, i.e., $\sum^N_{j=1} \cos(\frac{2 \pi j}{N}) = 0$. Though there is no  many body interactions in each of the systems, but coupling introduces an inter system many body interaction effect. This will generate inter-system correlation along with intra-system correlation.
We can construct the full Hamiltonian that governs the the dynamics of the joint quantum system from Eq.~\ref{eq:hamil_a} and Eq.~\ref{eq:hamil_ab} as the following, 
\begin{eqnarray}
&H(t)=\frac{1}{4}(\sum^N_{j_A=1}({\sigma}^x_{j_A}{\sigma}^x_{j_A+1}+{\sigma}^y_{j_A}{\sigma}^y_{j_A+1})\nonumber\\&+\sum^N_{j_B= 1}({\sigma}^x_{j_B}{\sigma}^x_{j_B+1}+{\sigma}^y_{j_B}{\sigma}^y_{j_B+1}))\nonumber\\&+ \frac{1}{2}(g \sum^N_{j_A= 1} \cos(\frac{2\pi j_A}{N})\sigma^z_{j_A} + g \sum^N_{j_B= 1} \cos(\frac{2\pi j_B}{N}) \sigma^z_{j_B} \nonumber \\&+ \varepsilon \sum^N_{j_A,j_B=1} \cos(\frac{2 \pi j_A}{N}) \cos(\frac{2 \pi j_B}{N})\sigma^z_{j_A}\sigma^z_{j_B} ) \sum_{n=- \infty}^{\infty} \delta(\frac{2\pi t}{\tau}-n).\nonumber\\
\end{eqnarray}

The operators ${U}$ and ${V}$ we introduced in Eq. \ref{eq:uv} are lattice translation and momentum translation operators respectively, and this can easily seen as: 
\begin{equation}
{V}|l \rangle = |l+1\rangle, \hspace{1cm}  \langle k|{U} = \langle k+1 |.
\end{equation}
If we replace ${V}$ and ${U}$ by discrete versions of standard quantum position and momentum translation operators $\exp(-i{p}a/\hbar)$ and $\exp(-i{x}b/\hbar)$ respectively Eq. \ref{eq:inte_hamil_t} leads to,
\begin{equation}
{H}(t) = \cos(\frac{{p}a}{\hbar}) + g\cos(\frac{{x}b}{\hbar}) \sum^{\infty}_{n =-\infty} \delta(\frac{2\pi t }{\tau} -n)
\end{equation}
Setting $a = b = 1/N$, $\hbar = 1/2\pi N$, and replacing the operators ${x}$ and ${p}$ by their classical values $x$ and $p$ respectively, the classical Harper Hamiltonian can be obtained as,
\begin{equation}
\label{eq:class_hamil}
H_c(t) = \cos(2\pi p) + g\cos(2\pi x) \sum^{\infty}_{n =-\infty} \delta(\frac{2\pi t }{\tau} -n),
\end{equation}
where the subscript `$c$' abbreviates the classical counterpart of the Hamiltonian. Similarly, the classical coupling Hamiltonian in Eq.~\ref{eq:hamil_ab} turns out to be,
\begin{equation}
\label{eq:class_hamil_ab}
\textcolor{black}{H^{AB}_c(t) = 2\varepsilon \cos(2\pi x^A) \cos(2\pi x^B)\sum^{\infty}_{n =-\infty} \delta(\frac{2\pi t }{\tau}-n).}
\end{equation} 
\textcolor{black}{The dynamics of the full system is governed by the Hamiltonians $H^A(t)$, $H^B(t)$ of the subsystems $A$ and $B$ and the interacting term $H^{AB}(t)$. The full Hamiltonian $H_c(t)$ is given by,
	\begin{eqnarray}
	\label{eq:class_hamil_coupled}
	H_c(t) &=& H^{A}_c(t) + H^{B}_c(t) + H^{AB}_c(t)\nonumber\\ &=&  \cos(2\pi p^A) + \cos(2\pi p^B)  + g(\cos(2\pi x^A) +  g\cos(2\pi x^B) \nonumber\\& & + 2\varepsilon \cos(2\pi x^A) \cos(2\pi x^B))  \sum^{\infty}_{n =-\infty} \delta(\frac{2\pi t }{\tau}-n). 
	\end{eqnarray}
}
\section{Observation of measure synchronization}
Now, since we have the Hamiltonian in the classical limit, we want to study here the dynamics of two coupled Harper systems and try to observe whether we have some analogous results in the quantum scenario. Since $\tau$ is the time interval between two consecutive kicks, then time $t = n \tau$, where $n$ is an integer. Following~\citep{lakshminarayan03}, the equation of motions of the coupled systems become:
\textcolor{black}{
	\begin{subequations}
		\begin{eqnarray}
		x^A(n+1)&=& x^A(n) -\tau \sin(2\pi p^A (n)), \\
		p^A(n+1)&=& p^A(n) +\tau g \sin(2\pi x^A (n+1)) \nonumber\\&&+ 2\tau \varepsilon \sin(2\pi x^A (n)) \cos(2\pi x^B (n)),\\
		x^B(n+1)&=& x^B(n) -\tau \sin(2\pi p^B (n)),\\
		p^B(n+1)&=& p^B(n) +\tau g \sin(2\pi x^B (n+1)) \nonumber\\&&+ 2\tau \varepsilon \sin(2\pi x^B (n)) \cos(2\pi x^A (n)).
		\end{eqnarray}
		\label{eq:EOM_classical}
	\end{subequations}
	The newly constructed maps (Eq.~\ref{eq:EOM_classical}) are defined within $0$ and $1$ using the standard `mod' function~\citep{lakshminarayan03}.} The dynamics of the individual system (i.e., with $H^{AB}_c = 0$) is already studied for different $\tau$ in~\citep{lakshminarayan03}. The individual system (i.e., either $H_c^{A}$ or $H_c^{B}$ with $H_c^{AB} = 0$) shows non-chaotic and chaotic motions at $\tau = 0.1$ and $\tau = 0.5$ respectively, and a mixed state of chaotic and non-chaotic dynamics is observed at $\tau = 0.3$~\citep{lakshminarayan03}. However, in this paper, we study the measure synchronization~\citep{hamp99} the coupled systems (i.e., with $H^{AB}_c \neq 0$) at different values of $\tau$. In the case of coupled Hamiltonian systems, we generally observed measure synchronization~\citep{hamp99, wang02, wang03, vincent05, gupta17}, which imply that in synchronized state the coupled systems share the identical regions (or, area) of the projected phase space---hence the name `measure'. In 
literature, there are two techniques, viz., 
equal average energy of participating systems~\citep{wang03, vincent05} and kink in average interaction energy~\citep{
	wang03} to detect the measure synchronized state. But, recently, two works~\citep{gupta17, ghosh18_2} reported that sometimes the equal bare energy technique fails to detect the synchronized state and encourage to use the other technique: kink in average interaction energy. Anyway, one can always back to the first principle of measure synchronization and compare the joint probability density functions of the coupled systems for the verification of the synchronized state~\citep{ghosh18_2}. However, for the further analysis, we adopt the joint probability density technique and the average interaction energy method to illustrate our results. From now onwards we use the term `synchronization' to refer the measure synchronization.
\begin{figure}[h]
	\includegraphics[width= 8.6 cm,height= 12 cm, keepaspectratio]{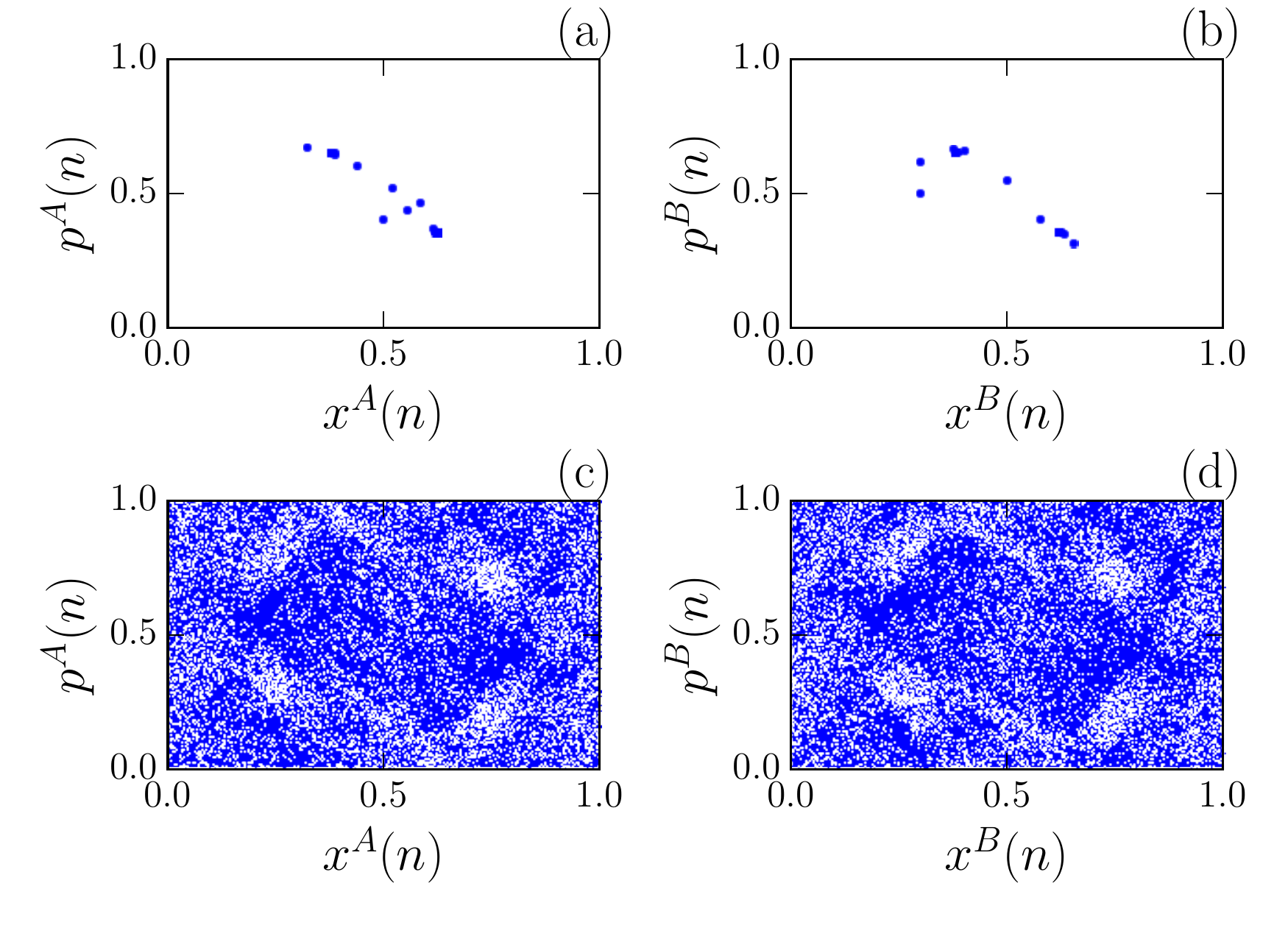}
	\caption{Measure desynchronization and measure synchronization are observed for different coupling strengths. The phase portraits of the individual subsystems, following Eq.~\ref{eq:EOM_classical}, are plotted for two different values of coupling strength $\varepsilon$ keeping fixed $\tau= 0.3$ and $g=1$ in both the cases. First row (a and b) is for $\varepsilon = 0.3$, whereas the last row (c and d) is for $\varepsilon = 0.7$. For $\varepsilon = 0.7$, both the subsystems are in synchronized state unlike the scenario at $\varepsilon = 0.3$. \textcolor{black}{In both the subplots, we have taken the initial condition as $(x^{A}(0), p^{A}(0), x^{B}(0), p^{B}(0)) = (0.5, 0.4, 0.3, 0.5)$.}}
	\label{fig:classical}
\end{figure}
\subsection{Joint probability density technique}
In Fig.~\ref{fig:classical}, we have studied the $\tau = 0.3$ case in detailed in presence of two different values coupling strengths $\varepsilon: 0.3$ (a and b) and $0.7$ (c and d), and observe that two coupled oscillators are not in the synchronized state at $\varepsilon = 0.3$, but in synchrony at $\varepsilon = 0.7$. Thus, the visual observation is a qualitative support of the synchronized state at $\varepsilon = 0.7$. Now before comparing the joint probability distribution function, let us briefly discuss the algorithm to calculate the joint probability distribution in the next paragraph.
The phase portrait of either system is bounded within $0$ and $1$---both in abscissa and ordinate. Let us divide the projected phase portrait---the $x^A-p^A$ plane (or, the $x^B-p^B$ plane)---into small square of area $\Delta x \Delta p= 1/M^2$, i.e., we divide either of the axes into $M$ small pieces with $\Delta x:=1/M$ and $\Delta p:=1/M$. Thus, $\rho^A (x^A, p^A) \Delta x \Delta p$ provides the fraction of total points within a square of area $\Delta x \Delta p$ with centre at $(x^A, p^A)$, where $\rho^A (x^A, p^A)$ is the joint probability distribution of the subsystem $A$. Similarly, we can define the joint probability distribution function---$\rho^B (x^B, p^B)$---of the second subsystem $B$.  Now, the coupled subsystems are in measure synchronized state if $\Delta \rho:= |\rho^A (x, p)-\rho^B (x, p)| \leq \rho_c \, \forall \, (x, p)$, where $\rho_c$ is a threshold and should be infinitesimally small.   
\begin{figure}[h]
	\includegraphics[width= 8.6 cm,height= 12 cm, keepaspectratio]{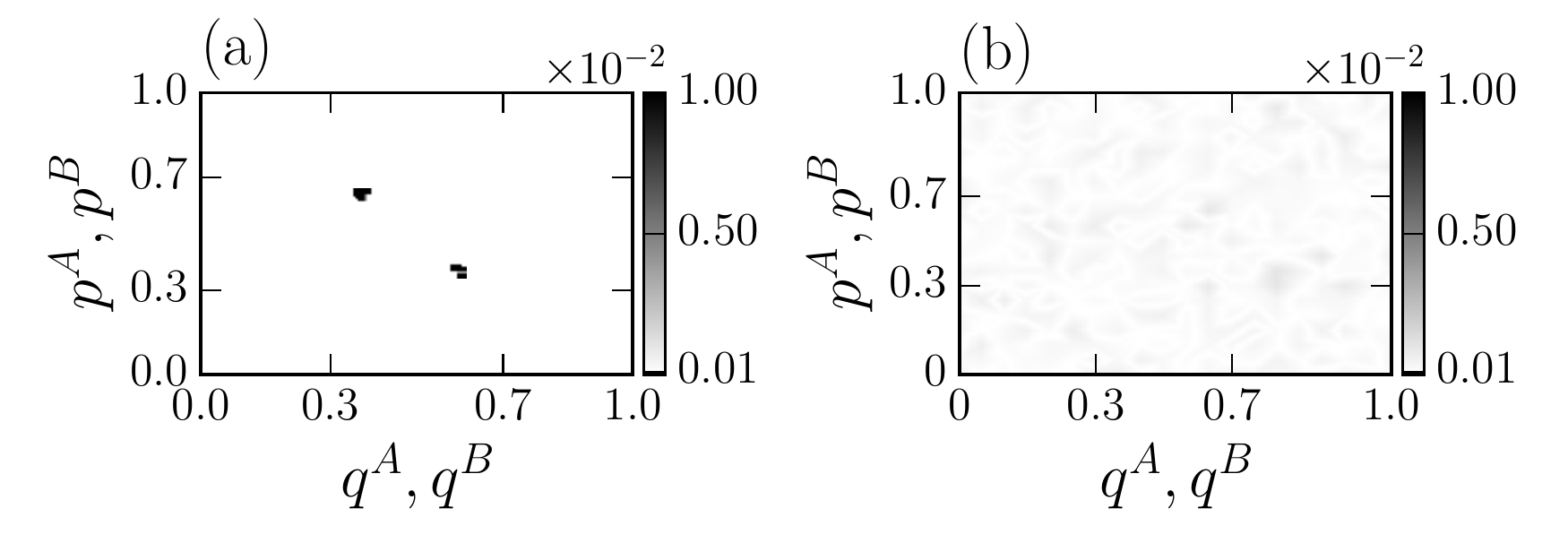}
	\caption{Fig.~\ref{fig:classical} is depicted here from a quantitative point of view: comparing the joint probability densities of the coupling subsystems. $\Delta \rho$ is plotted in colour in (a) and (b) for $\varepsilon = 0.3$ and $0.7$ respectively. In (a), $\Delta \rho$ is larger than the threshold $\rho_c$ ($=1\times 10^{-3}$), indicating the desynchronized state. But in (b), the condition $\Delta \rho \leq \rho_c$ is always satisfied indicating the synchronized state.}
	\label{fig:classical_jpd}
\end{figure}
In theory, $\rho_c$ should be zero. However, this demands that the time-series under study be infinitely long and be sampled continuously in time. In practice, this is impossible. Hence, the best we can do is to choose $\rho_c$ to be some small non-zero number depending on how long the systems are evolved in the numerical simulations. We have taken $M = 20$, i.e., the joint probability distribution is calculated over total $20 \times 20$ grids, and the threshold $\rho_c$ is taken as $1\times 10^{-3}$. Note that there is nothing special about $M = 20$. One may choose any larger $M$. The choice of smaller values of $M$ has the obvious problem that the information of the local dynamics can’t be captured---as an extreme example, $M=1$ can only capture an averaged global measure. Also, in practice, $\rho_c$ depends on the $M$ value---this dependence would also vanish in the limit of infinitely long data that has been sampled continuously.
\subsection{Average interaction energy technique}
\begin{figure}[h]
	\includegraphics[width= 8.6 cm,height= 12 cm, keepaspectratio]{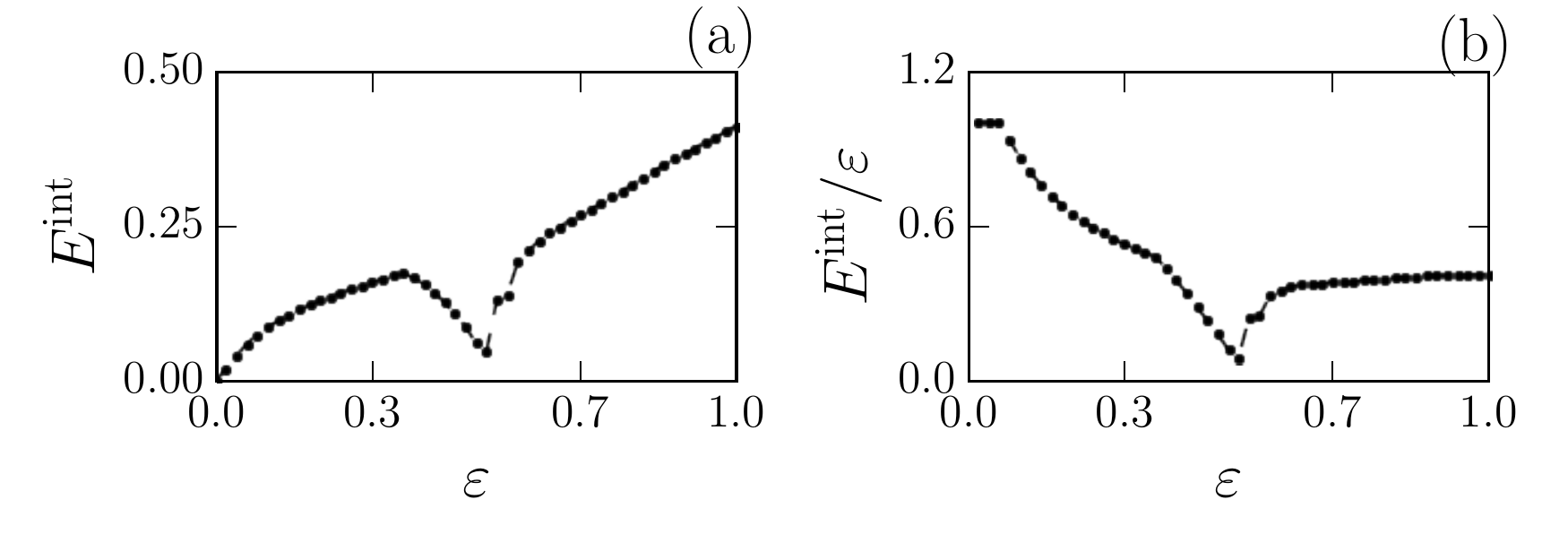}
	\caption{The average interaction energy $E^{\rm int}$ is plotted for the coupled subsystems $A$ and $B$ following Eq.~\ref{eq:EOM_classical} and \ref{eq:int_energy} for different coupling strengths $\varepsilon$ \textcolor{black}{with the initial condition $(x^{A}(0), p^{A}(0), x^{B}(0), p^{B}(0)) = (0.5, 0.4, 0.3, 0.5)$}. A kink observed at $\varepsilon = 0.5$ shows the transition from the desynchronized state to the synchronized state.}
	\label{fig:classical_int_en}
\end{figure}

\begin{figure}[h]
	\includegraphics[width= 8.6 cm,height= 12 cm, keepaspectratio]{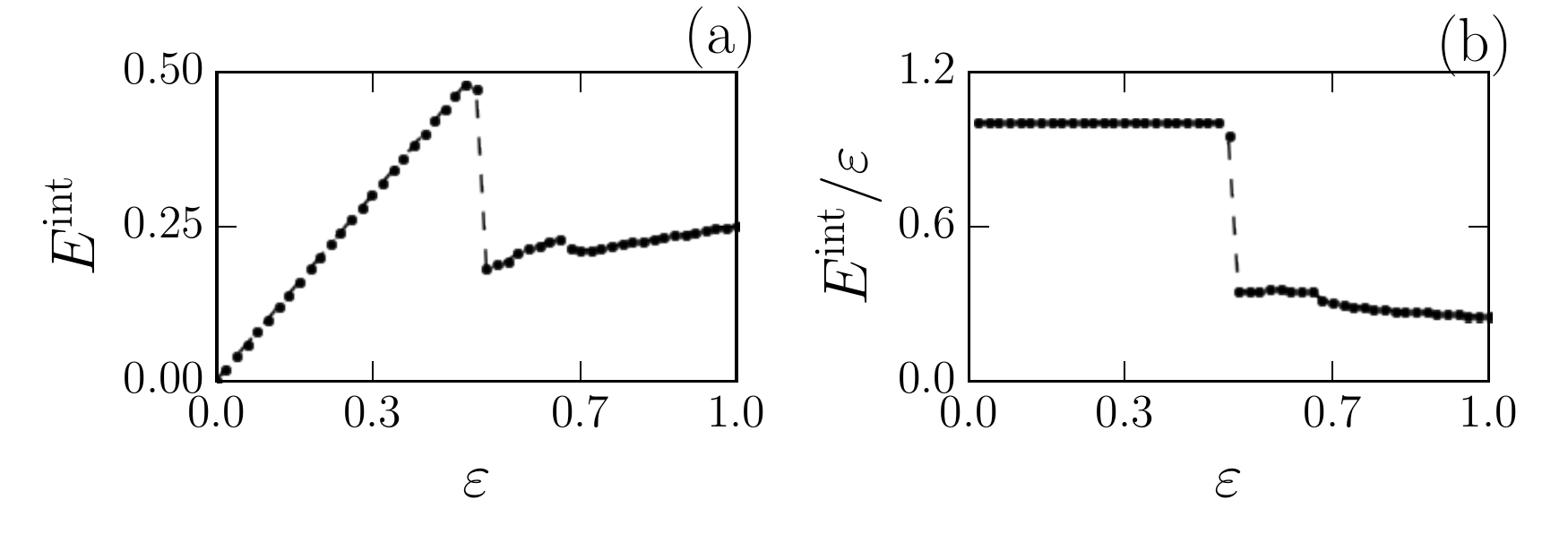}
	\includegraphics[width= 8.6 cm,height= 12 cm, keepaspectratio]{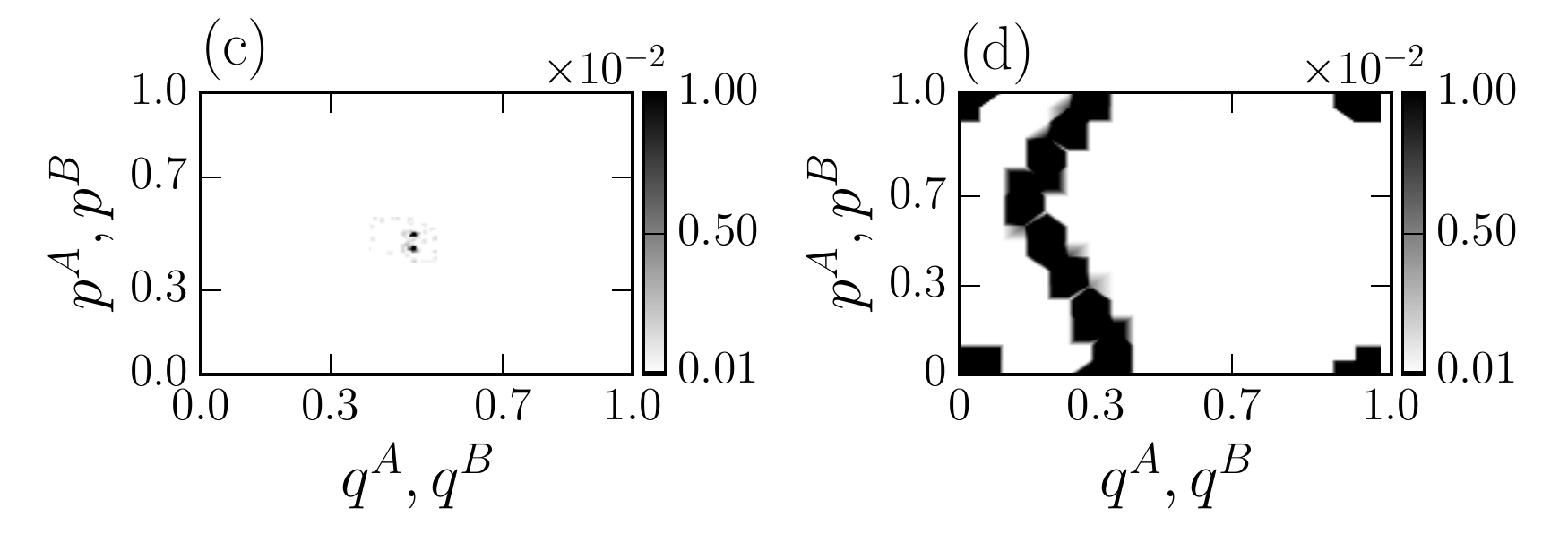}
	\caption{Similar to Fig.~\ref{fig:classical_int_en} and \ref{fig:classical_jpd},  $E^{\rm int}$ and $\Delta \rho$ are plotted for the coupled subsystems $A$ and $B$ for different coupling strengths $\varepsilon$ with $\tau = 0.1$ and \textcolor{black}{unaltered initial conditions $(x^{A}(0), p^{A}(0), x^{B}(0), p^{B}(0)) = (0.5, 0.4, 0.3, 0.5)$}. A discontinuity observed at $\varepsilon = 0.5$, but no transition is observed at $\varepsilon = 0.5$. Subplots (c) and (d) show the variation of $\Delta \rho$ for $\varepsilon= 0.3$ and $0.7$ respectively. The higher values of $\Delta \rho$ ($\Delta \rho > \rho_c$ with $\rho_c=4\times 10^{-4}$)  imply the desynchronized state in both the cases.}
	\label{fig:classical_tau_p1}
\end{figure}

\begin{figure}[h]
	\includegraphics[width= 8.6 cm,height= 12 cm, keepaspectratio]{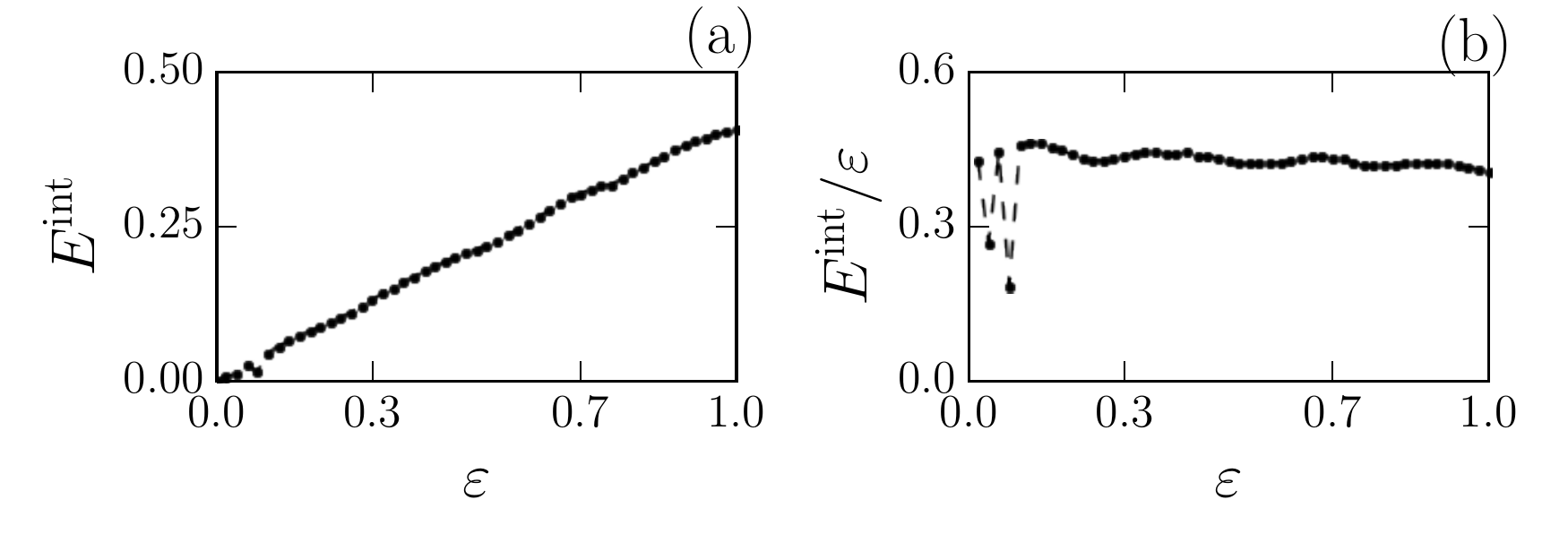}
	\includegraphics[width= 8.6 cm,height= 12 cm, keepaspectratio]{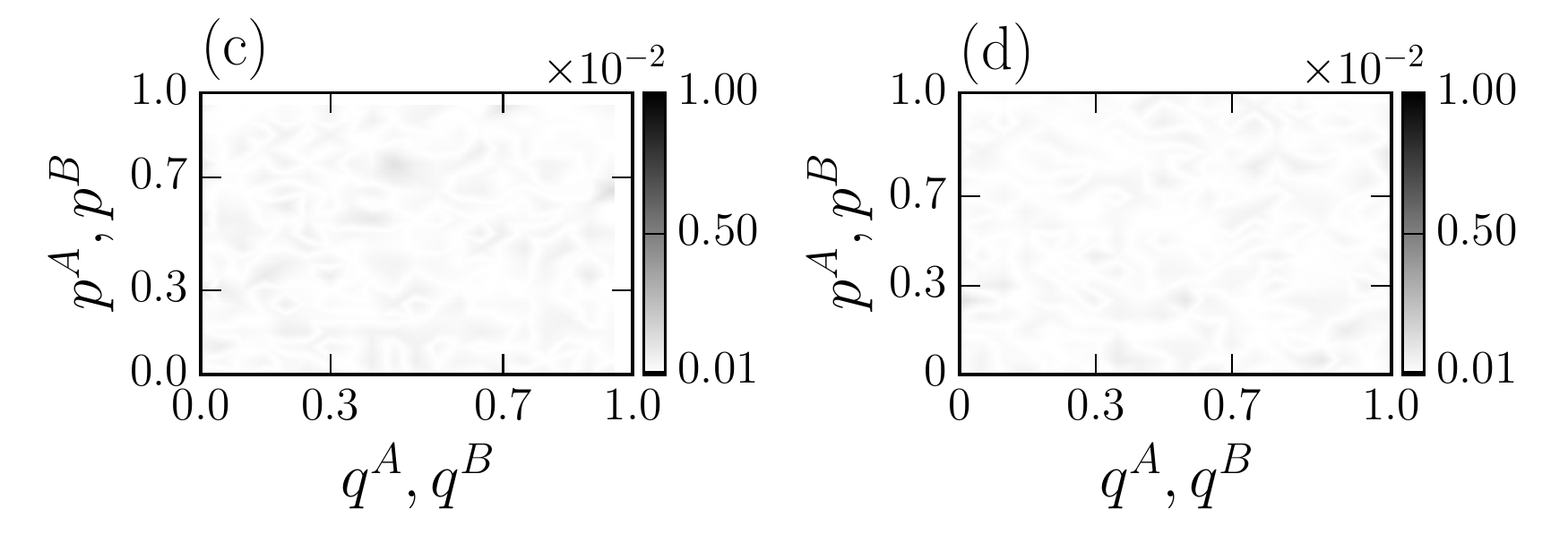}
	\caption{This plot is for $\tau = 0.5$ \textcolor{black}{with the initial condition $(x^{A}(0), p^{A}(0), x^{B}(0), p^{B}(0)) = (0.5, 0.4, 0.3, 0.5)$}. Here no kink is observed in $E^{\rm int}$ for different coupling strengths $\varepsilon$, which supports the absence of the transition. The density plots of $\Delta \rho$, for $\varepsilon = 0.3$ (subplot (c)) and $0.7$(subplot (d)), are also in accordance with this conclusion.}
	\label{fig:classical_tau_p5}
\end{figure}

As discussed at the beginning of this section, the average interaction energy indicate the transition from a synchronized state to desynchronized state or vice versa. Here, the average interaction energy of the coupled subsystems ($A$ and $B$), following Eq.~\ref{eq:class_hamil_coupled}, is given by: 
\begin{equation}
E^{\rm int} = \frac{1}{T} \sum_{n = 1}^{T} H^{AB}_{c} (n).
\label{eq:int_energy}
\end{equation}
In addition with the phase coordinates, $ E^{\rm int}$ depends explicitly on the the coupling strength. The average interaction energy is plotted in Fig.~\ref{fig:classical_int_en}(a) with different $\varepsilon$ using Eq.~\ref{eq:EOM_classical} and \ref{eq:int_energy}. One kink is observed at $\varepsilon = 0.5$, which further implies that there is a transition from the desynchronized state to the synchronized state at $\varepsilon = 0.5$~\citep{gupta17} as $\varepsilon$ is increased. For more clear understanding, we have plotted the phase portraits at $\varepsilon = 0.3$ and $0.7$ in Fig.~\ref{fig:classical}. Fig.~\ref{fig:classical} leads to the conclusion that at $\varepsilon = 0.3$, the participating subsystems are in desynchronized state, whereas at $\varepsilon = 0.7$ they occupy the same measure in the projected phase space indicating the synchronized state. 
Thus we may conclude here, with the help of above mentioned techniques, that synchronized state is observed for coupled Harper systems after $\varepsilon = 0.5$ using $\tau = 0.3$. However, the scenario is different for other values of $\tau$. For example, if we choose $\tau = 0.1$, we observe a discontinuity in $E^{\rm int}$ at $\varepsilon = 0.5$ (see Fig.~\ref{fig:classical_tau_p1}(a) or (b)), and following 
our discussion there should be a transition between the desynchronized and the synchronized state. But, the variation of $\Delta \rho$ (see Fig.~\ref{fig:classical_tau_p1}(c) or (d)) explain that for both the cases: $\varepsilon = 0.3$ and $0.7$, desynchronized state is observed. Thus, the discontinuity in $E^{\rm int}$, observed here, misguided us to make a decision, and the aforesaid problem continued for all values of $\tau \lesssim 0.3$. Besides, for $\tau \gtrsim 0.3$, we do not observe the transition for any $\varepsilon \in [0, 1]$. In Fig.~\ref{fig:classical_tau_p5} we have explained the case for $\tau = 0.5$ and no kink is observed in $E^{\rm int}$ (see Fig.~\ref{fig:classical_tau_p5}(a) or (b)), which maybe because of the full-fledged chaotic nature of the coupled subsystems at $\tau = 0.5$~\citep{lakshminarayan03, lima91} and they are always in synchronized state. The values of $\Delta \rho$ in Fig.~\ref{fig:classical_tau_p5}(c) or (d) support the existence of the synchronized states for $\
varepsilon = 0.3$ and $0.7$. This 
further help us to make a possible conclusion that \emph{the transition is always associated with a kink in $E^{\rm int}$, but any non-analyticity in $E^{\rm int}$ does not imply the transition}. \textcolor{black}{However, all the conclusions drawn from the results in this section remain unchanged from other sets of initial conditions also.}
\section{Quantum Dynamics}
Now, we turn our discussions on time evolution of quantum systems $A$ and $B$, and each of the systems is an $N$ qubit system. The operators in  the systems $A$ and $B$ belong to disjoint Hilbert spaces $\mathbb{H}^A$ and $\mathbb{H}^B$ respectively. Even initially each of the system starts from a pure state, due to the coupling eventually they will entangle with each other and become mixed states hence we have to evolve both the systems simultaneously. Let $|\psi^A_0\rangle$ and $|\psi^B_0\rangle$ be the initial states of systems $A$ and $B$ respectively. The state of the joint system can be written as $|\psi_0\rangle = |\psi^A_0\rangle \otimes |\psi^B_0\rangle$. We consider evolution at discrete times, viz., $t=\tau^+ /2\pi ,2\tau^+ /2\pi$, etc., i.e., at time instant just after a kick. The state of the joint system after $n$ kicks will be given by:
\begin{eqnarray}
|\psi_{n\tau/2 \pi}\rangle = \mathcal{U}^n|\psi_{0}\rangle = \mathcal{U}^n|\psi^A_0 \rangle \otimes |\psi^B_0\rangle,
\end{eqnarray}
where, the time evolution operator $\mathcal{U}$ is defined on the joint Hilbert space $\mathbb{H}^A \otimes \mathbb{H}^B$  just for one kick (corresponding to $t=\tau/2\pi$).
To solve the system analytically, we need to focus on the symmetry of the dynamics. It can be easily seen that number of down spins is conserved throughout the dynamics in the joint system as well as in the individual systems. The XY dynamics can be studied analytically using Bethe Ansatz eigenfunctions \citep{bethe}. The one-magnon excitations can be created by turning any one of the spins, giving $N$ localized one-magnon states. One-magnon eigenstates are labelled by the momentum of the down spin and the corresponding eigenfunction is given by:
\begin{eqnarray}
\phi_p^x = & \sqrt{\frac{1}{N}} e^{i px}; p ={ 2\pi I\over N}, {\rm for ~a~ closed ~chain}, 
\end{eqnarray}
where the momentum $p$ is determined by an integer $I=1,2,...,N$. The one-magnon eigenvalue is given by $\epsilon_1(p)=-\cos{p}$. The time evolution of the system will transport the single down spin from site $x$ to another site $x'$ and the time dependent probability is given by one magnon Green function, where the time-dependent function $G^{x'}_{x}(t)$~\citep{vs04,sur17} is given by:
\begin{equation} 
G^{x'}_x(t)  = (-i)^{x-x'}J_{x-x'}(t) ,
\end{equation}
for a closed chain. $J_x(y)$ is the Bessel function of integer order $x$ and argument $y$. The system evolves between a time $n\tau^+ /2\pi$ to $(n+1)\tau^- /2\pi$ through XY dynamics between two kicks which introduces a lattice position dependent phase factor to the Green function~\citep{sur18_2}. It can be seen that after each kick, a site-dependent new phase is introduced in the Green function which indicates the qualitative change in the dynamics from XY dynamics, i.e., the state after a kick depends on  the location of the down spin after the previous kick. 
The most general initial state to start the dynamics with is $M_A$ number of down spins in system $A$ and $M_B$ number of down spins in system $B$. The most trivial state to start the dynamics is where there is no down spin in one of the systems. But for such initial state the systems will not get correlated through the coupling. \textcolor{black}{Since no bound state is formed even for multiple number of down spins in the individual systems, the wave function for any general initial state can be obtained from the product of non-interacting one particle Green functions. So, no new physics is coming out for any multimagnon state as the initial state. But to observe the transition, it is necessary that the individual systems start with different initial states. Since we have discussed that the dynamics is qualitatively independent of the initial conditions, any initial states is sufficient to observe the transition. Hence, we restrict ourselves to one particle initial states ($M_A = 1$ and $M_B =1$, i.e., a 
direct-product state with $1+1$ number of down spins) for simplicity of calculations}. The initial state is denoted as:
\begin{equation}
\label{eq:intial_st}
|\psi_0\rangle = |x_0;y_0\rangle,
\end{equation}
where $x_0(y_0)$ denotes the location of the down spin in system $A(B)$. $x_i$ are the co-ordinates of the sites with down spins in system $A$ after $i^{th}$ kick, and similarly,  $y_i$ are for system $B$.
\textcolor{black}{We show the time evolution of the state explicitly for one kick. The time evolution operator $\mathcal{U}$ acting on the the state $|x_0;y_0\rangle$ yields the following state at time $ t = \tau/2 \pi$,}

		\begin{eqnarray}
		\label{eq:one_kick}
		&|\psi_{t = \tau/2 \pi}\rangle = \mathcal{U} |x_0;y_0\rangle \nonumber\\ 
		&=  e^{\frac{i}{2}g (\sum^N_{j_A = 1} \cos(\frac{2\pi j_A}{N})\sigma^z_{j_A} + \sum^N_{j_B= 1} \cos(\frac{2\pi j_B}{N}) \sigma^z_{j_B} )} \nonumber\\ &\times e^{\frac{i}{2}\varepsilon \sum^N_{j_A,j_B=1} \cos(\frac{2 \pi j_A}{N}) \cos(\frac{2 \pi j_B}{N})\sigma^z_{j_A}\sigma^z_{j_B} } \nonumber \\ & \times e^{\frac{i\tau}{4} \left(\sum^N_{j_A= 1}({\sigma}^x_{j_A}{\sigma}^x_{j_A+1}+{\sigma}^y_{j_A}{\sigma}^y_{j_A+1})+\sum^N_{j_B=1}({\sigma}^x_{j_B}{\sigma}^x_{j_B+1}+{\sigma}^y_{j_B}{\sigma}^y_{j_B+1})\right)} |x_0;y_0\rangle \nonumber\\ &= \sum_{x_1,y_1} {G}^{x_1}_{x_0}(\frac{\tau}{2 \pi}) {G}^{y_1}_{y_0}(\frac{\tau}{2 \pi}) e^{i \left(g\cos(\frac{2 \pi {x_{1}}}{N})+g\cos(\frac{2 \pi y_{1}}{N}) -2\varepsilon \cos(\frac{2 \pi x_{1}}{N})\cos(\frac{2 \pi y_{1}}{N})\right)}\nonumber\\&|x_1;y_1\rangle  
		\end{eqnarray}

	\textcolor{black}{Extending Eq.~\ref{eq:one_kick} by recursion relation for $n$ kicks the joint quantum state at time $t = n\tau/2 \pi$ is given in following form  :}
	\begin{eqnarray}
	|\psi_{t = n\tau/2 \pi}\rangle = \sum_{x_n,y_n} \tilde{G}^{x_n;y_n}_{x_0;y_0}(n)|x_n;y_n\rangle,
	\end{eqnarray}
	where the wave function $\tilde{G}^{x_n;y_n}_{x_0;y_0}(n)$ in the above equation is the Green function of the coupled joint system and given in terms of one magnon Green functions as:

		\begin{eqnarray}
		\label{eq:final_st}
		&\tilde{G}^{x_n;y_n}_{x_0;y_0}(n) = \sum_{x_1,x_2,...,x_{n-1}} \sum_{y_1,y_2,...y_{n-1}} \prod^{n-1}_{i=0}G^{x_{i+1}}_{x_i}(\frac{\tau}{2 \pi}) G^{y_{i+1}}_{y_i}(\frac{\tau}{2 \pi})\nonumber\\ &\times e^{i (g\cos(\frac{2 \pi {x_{i+1}}}{N})+g\cos(\frac{2 \pi y_{i+1}}{N}) -2\varepsilon \cos(\frac{2 \pi x_{i+1}}{N})\cos(\frac{2 \pi y_{i+1}}{N}))}. 
		\end{eqnarray} 

	Energies of the individuals systems $A$ and $B$ as well as the interaction energy can be computed from the above wave function. Energies $E^A$ and $E^B$ of the systems $A$ and $B$ respectively  after $n$ kicks are given by,
	
	\begin{eqnarray}
	E^A_{n} = \frac{1}{2}\sum_{x_n,y_n}  \tilde{G}^{* x_n;y_n}_{x_0;y_0}(n) (\tilde{G}^{x_n+1;y_n}_{x_0;y_0}(n)+\tilde{G}^{x_n-1;y_n}_{x_0;y_0}(n));\nonumber\\
	E^B_{n} = \frac{1}{2}\sum_{x_n,y_n}  \tilde{G}^{* x_n;y_n}_{x_0;y_0}(n) (\tilde{G}^{x_n;y_n+1}_{x_0;y_0}(n)+\tilde{G}^{x_n;y_n-1}_{x_0;y_0}(n)).\nonumber\\
	\end{eqnarray} 
	
	The interaction energy after $n$ kicks is given by,
	\begin{eqnarray}
	E^{\rm int}_{n} = \sum_{x_n,y_n} | \tilde{G}^{x_n;y_n}_{x_0;y_0}(n)|^2 \langle x_n;y_n |H^{AB}|x_n;y_n\rangle \nonumber\\
	=  2 \varepsilon \sum_{x_n,y_n} |\tilde{G}^{x_n;y_n}_{x_0;y_0}(n)|^2  \cos(\frac{2 \pi x_n}{N})\cos(\frac{2 \pi y_n}{N}).
	\end{eqnarray} 
	Unlike the classical scheme, where the interaction takes place only between two systems through the coupling, the quantum counterpart is more complicated. Here, the individual systems ($A$ and $B$) as well as individual qubits in each systems though uncorrelated initially, will generate multi party correlation through time evolution. $N$ qubits in system $A$ will get correlated with $N$ qubits in system $B$, this is a multi party correlation. Also each qubit in system $A$(or $B$) will get correlated with the rest $(N-1)$ qubits in system $A$ (or $B$) as well as $N$ qubits of system $B$ (or $A$). The correlation between system $A$ and $B$ is quantified through the reduced density matrix (RDM) of the systems. Since we know the joint state at any time $t$, the RDM for any system can be computed by tracing out the rest from the joint density matrix. The definition of the RDM for any subsystem $X$ is given by,
	\begin{eqnarray}
	\rho^{X} = Tr_{\bar{X}}~ |\psi\rangle\langle\psi|,
	\end{eqnarray}
	where, $Tr_{\bar{X}}$ is the partial trace over the Hilbert space excluding the subsystem $X$. All informational based correlation measures are computed from the RDM, e.g., the von Neumann  entropy of the subsystem $A$ will be given as,
	\begin{eqnarray}
	S^A_t  = - Tr~ \rho^{A}_t \ln \rho^{A}_t . 
	\end{eqnarray} 
	The von Neumann entropy $S^A_t$ measures entanglement between the system $A$ and the rest.  Also, there are other informational based measures for quantum correlations. The linear entropy for the subsystem $A$ is defined as,
	
	\begin{eqnarray}
	\mathcal{T}^A_t = 1 - Tr~ {\rho^{A}_t}^2. 
	\end{eqnarray}
	The concurrence\citep{wootter, rungta} between the individual systems $A$ and $B$ is given as,
	\begin{eqnarray}
	\mathcal{C}^{AB}_t = \sqrt{2(1 - Tr~ {\rho^{A}_t}^2)}. 
	\end{eqnarray}

	\begin{figure*}[t]
		\begin{center}
			\subfigure[]{%
				\label{fig:first}
				\includegraphics[width=0.26\textwidth]{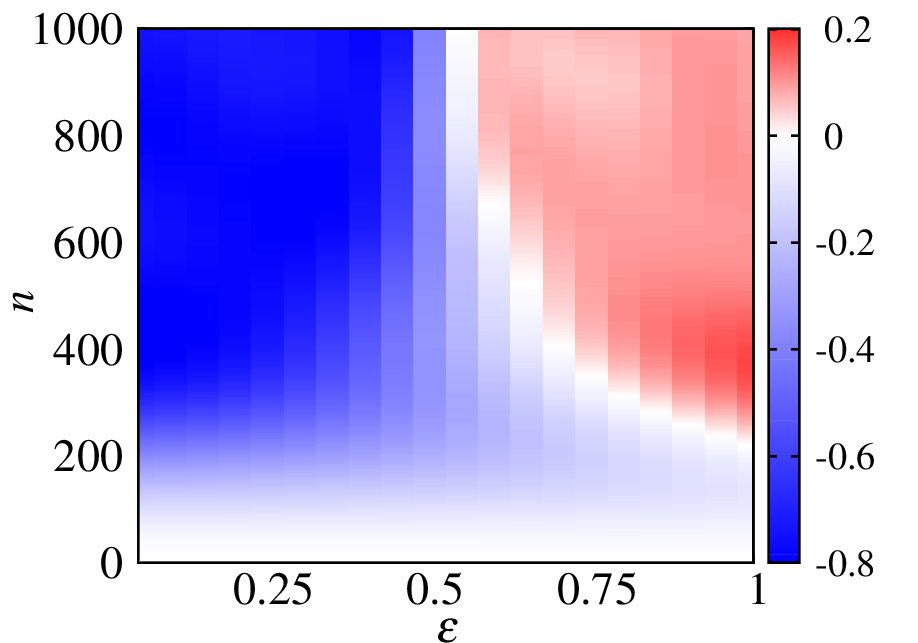}
			}%
			\subfigure[]{%
				\label{fig:second}
				\includegraphics[width=0.26\textwidth]{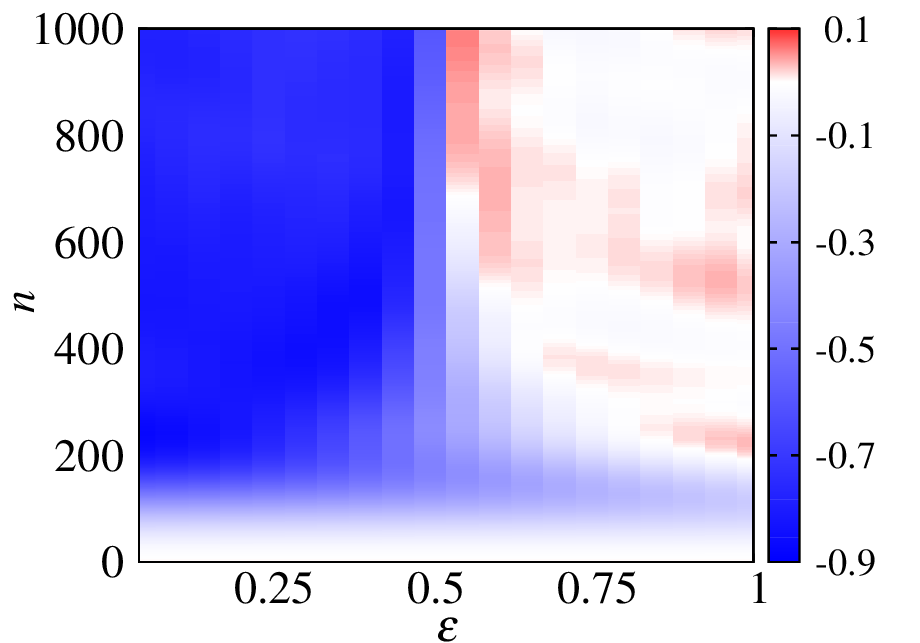}
			}%
			\subfigure[]{%
				\label{fig:first}
				\includegraphics[width=0.26\textwidth]{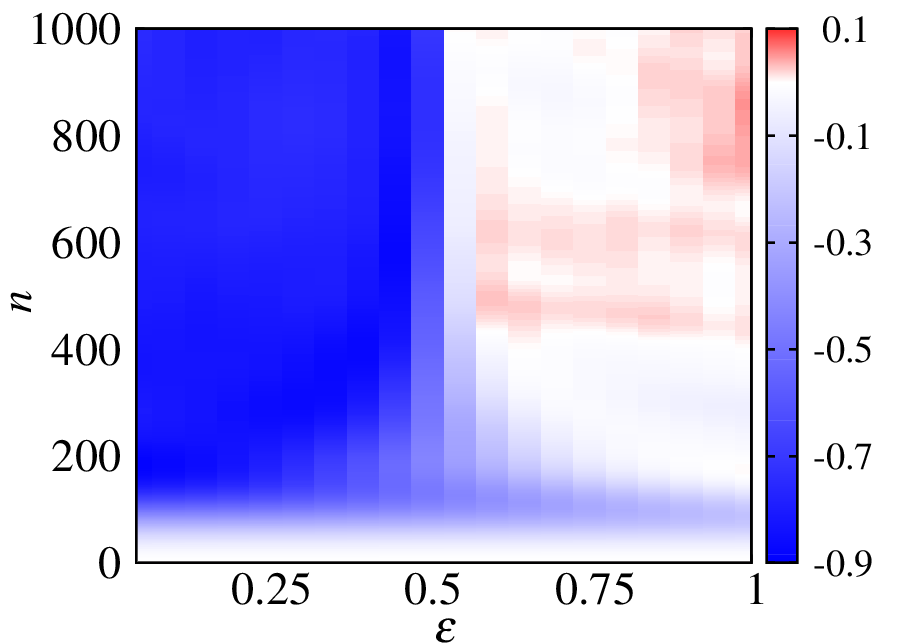}
			}\\%
			\subfigure[]{%
				\label{fig:second}
				\includegraphics[width=0.26\textwidth]{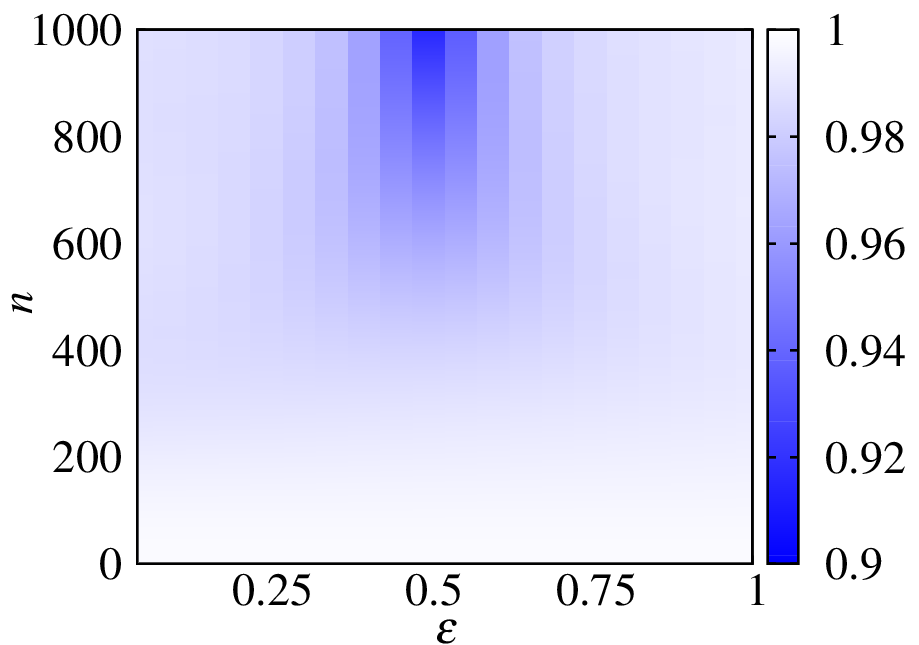}
			}%
			\subfigure[]{%
				\label{fig:second}
				\includegraphics[width=0.26\textwidth]{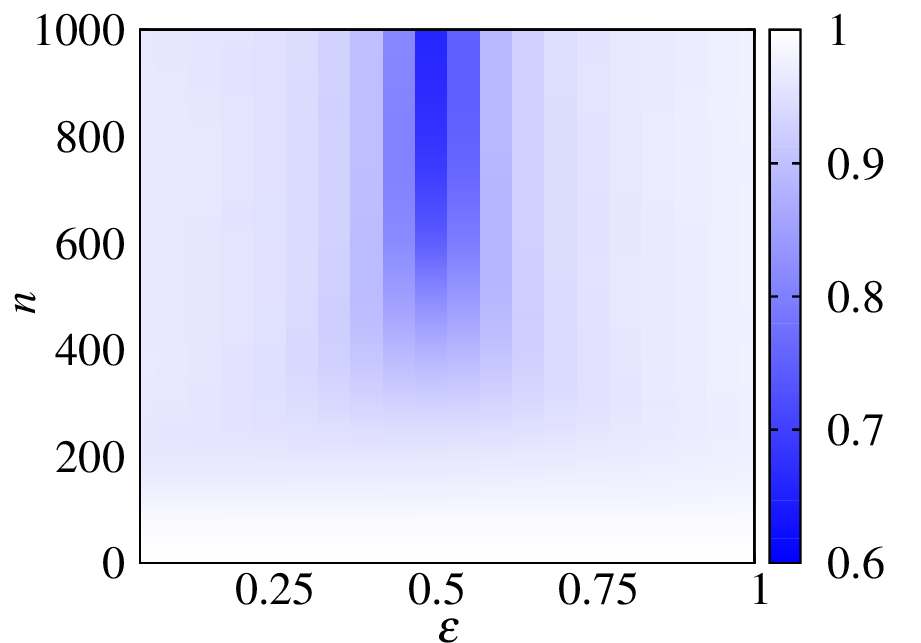}
			}%
			\subfigure[]{%
				\label{fig:second}
				\includegraphics[width=0.26\textwidth]{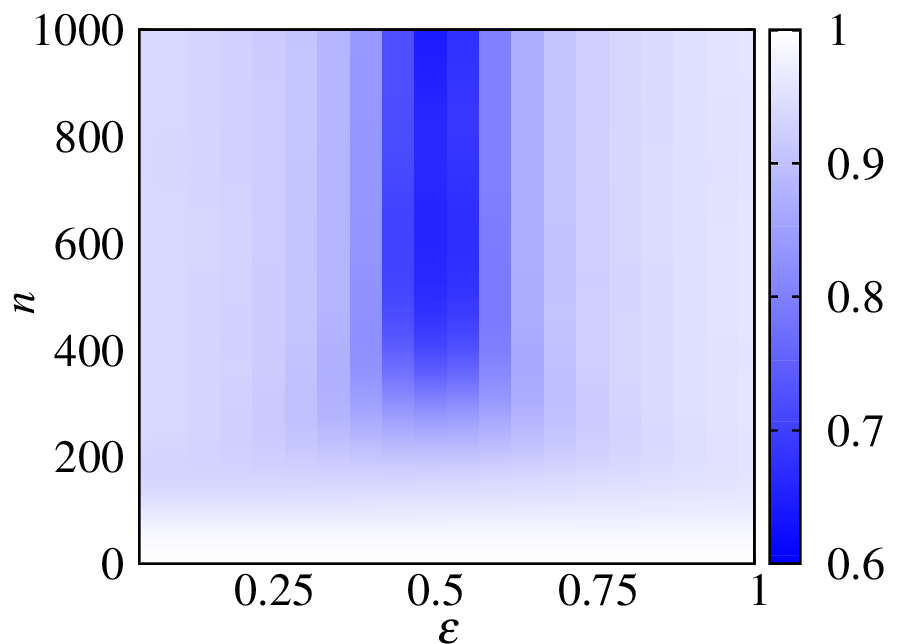}
			}\\
			\subfigure[]{%
				\label{fig:second}
				\includegraphics[width=0.26\textwidth]{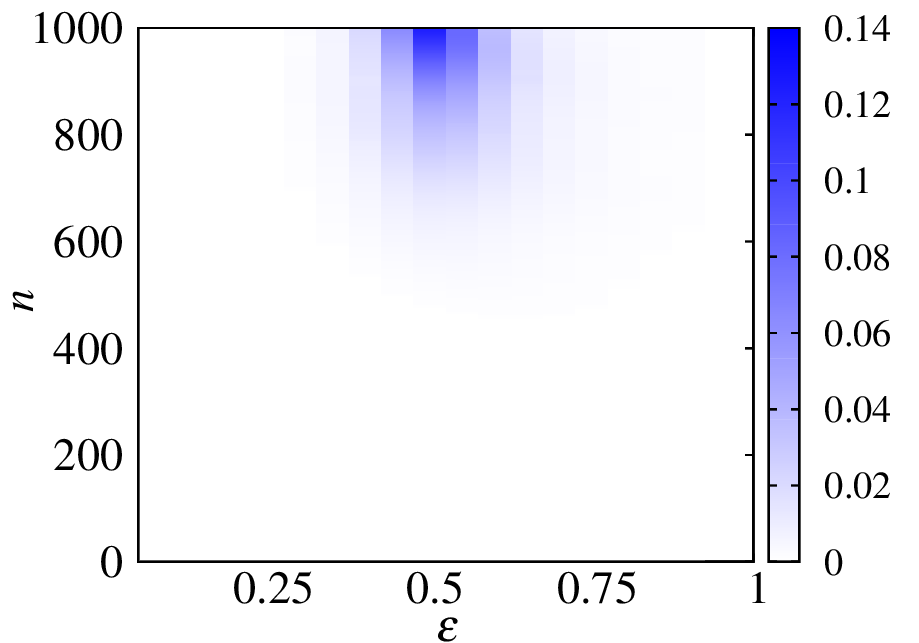}
			}%
			\subfigure[]{%
				\label{fig:second}
				\includegraphics[width=0.26\textwidth]{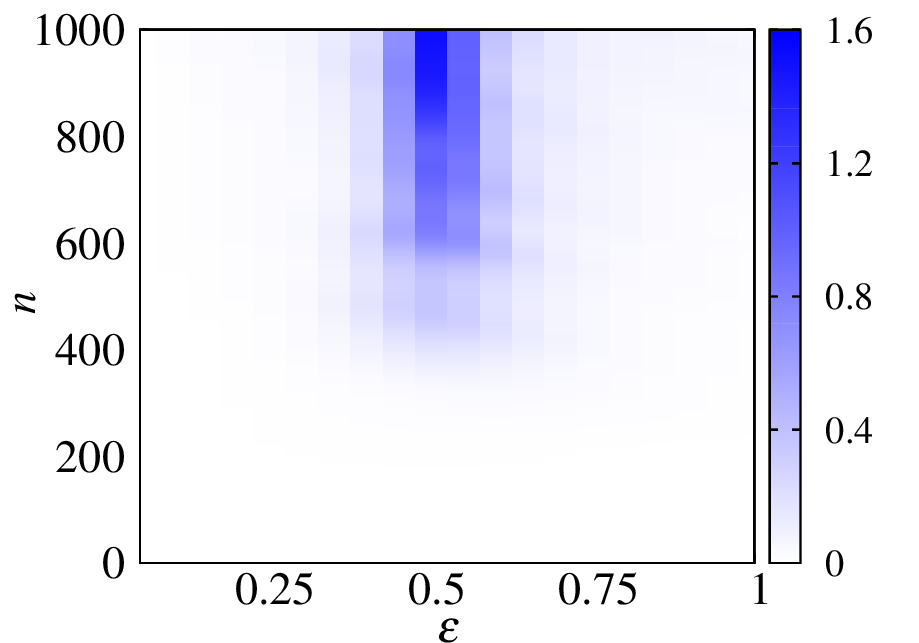}
			}%
			\subfigure[]{%
				\label{fig:second}
				\includegraphics[width=0.26\textwidth]{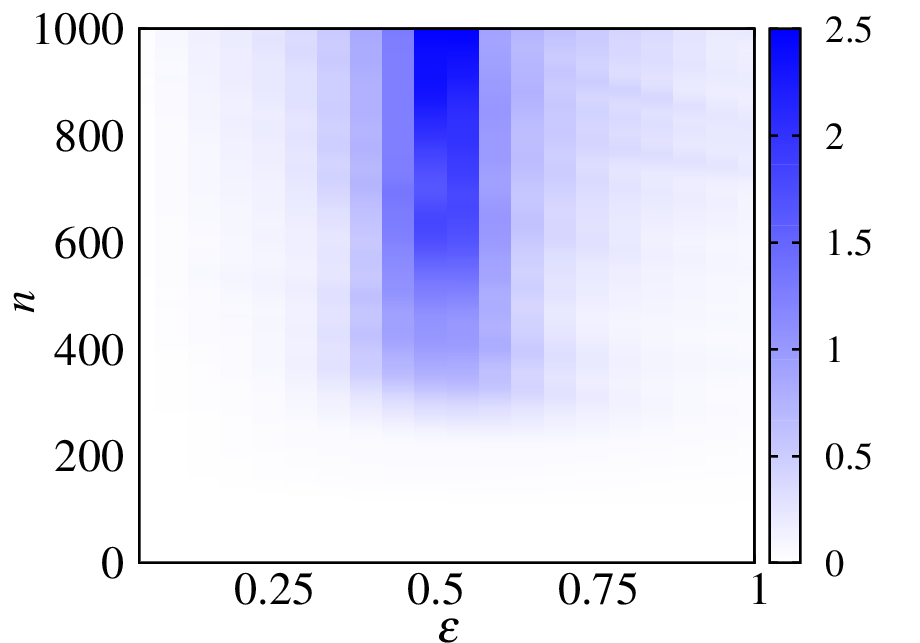}
			}\\%
		\end{center}
		\caption{\label{fig:fig_1}{\emph{(color online)} The difference between the energies of system $A$ and $B$ ($\Delta E = E^A - E^B$) as a function of number of kicks ($n$) and the coupling parameter ($\varepsilon$) for kicking intervals (a)$0.1/2\pi$ (b)$0.3/2\pi$  (c)$0.5/2\pi$. The average interaction energy divided by twice of the coupling parameter ($E^{\rm int}/2\varepsilon$) as a function of number of kicks ($n$) and the coupling parameter ($\varepsilon$) for kicking intervals (d)$0.1/2\pi$ (e)$0.3/2\pi$  (f)$0.5/2\pi$. The von Neumann entropy ($S^A$ or $S^B$) as a function of number of kicks ($n$) and the coupling parameter ($\varepsilon$) for kicking intervals (g)$ 0.1/2\pi$ (h)$0.3/2\pi$  (i)$0.5/2\pi$. Number of qubits in each system: $N = 100$. Initial state: $x_0 = 1$, $y_0 = N/2$.}
		}%
		\label{fig:subfigures}
		\label{fig:quantum}   
	\end{figure*}
	
	The linear entropy, the lowest order approximation of the von Neumann entropy quantifies the mixedness of a quantum state. While $\mathcal{C}^{AB}_t$ quantifies entanglement between the subsystems $A$ and $B$. All the three measures $S^A_t$, $\mathcal{T}^A_t$, and $\mathcal{C}^{AB}_t$ vanish for separable states.  Since the joint state here is a pure state the mutual information between systems $A$ and $B$ is just twice the von Neumann entropy of the either system. So it is sufficient to discuss only $S^A_t$  to describe the dynamics. Here, we have taken the system size $N = 100$ to illustrate our results. Note that the maximum value of von Neumann entropy corresponding to each system is $S^A_{\rm max} = \ln {100} \approx 4.6052$ for one particle states.
	Fig.~\ref{fig:quantum} show the time dependence of  difference between average energies of individual systems ($\Delta E = E^A -E^B$), the average interaction energy divided by twice coupling parameter, i.e., $E^{\rm int}/2\varepsilon$, between the systems $A$ and $B$ and the von Neumann entropy ($S^A$) of the systems $A$ or $B$  for different values of $\varepsilon$ and $n$. The initial state we have taken is $|\psi_0\rangle = |1,N/2\rangle$ and number of qubits in each system $N = 100$. The state $|\psi_0\rangle = |1,N/2\rangle$ means the down spin is at the first site in system $A$ and in site  $N/2$  in system $B$. Since the systems are closed or ring like, the down spins are farthest in this particular initial state and thus most unlikely to couple.
	We see that all the three quantities $\Delta E$, $E^{\rm int}/2\varepsilon$, and $S^A$ are non-monotonic functions of $\varepsilon$ after certain number of kicks. As shown in Fig.\ref{fig:quantum}(a)--(c) the difference between the average energies ($\Delta E$) of the individual systems $A$ and $B$ initially grows negative from zero with time, independent of the value  $\varepsilon$ for all values of $\tau$. As shown in Fig. \ref{fig:quantum}(a) $\Delta E$ becomes negative for $\varepsilon < 0.5$ and slightly positive for $\varepsilon > 0.5$ after certain value of $n$. This means starting from the same initial state with two different values of coupling constant ($\varepsilon < 0.5$ or, $\varepsilon > 0.5$), results in two completely different energy distributions in the systems $A$ and $B$. For  large $n$ and $\varepsilon > 0.5$, the quantity $\Delta E$  is almost zero for $\tau = 0.3$ and $\tau = 0.5$ as shown in Fig. \ref{fig:quantum}(b) and (c) respectively. This indicates that the energies of the 
individual systems tend to have a common value  for $\varepsilon > 0.5$ and different for $\varepsilon < 0.5$ for kicking period parameter $\tau \gtrsim 0.3$. It is a clear indication of a dynamical phase transition at $\varepsilon = 0.5$. So, the quantity $\Delta E$ may serve as an order parameter for the transition.
	Since the systems start from uncoupled state the average interaction energy is zero initially. As time evolves they get coupled and develops nonzero average interaction energy  and $E^{\rm int}/2\varepsilon$ shows a minimum at $\varepsilon = 0.5$ for all values $\tau$, as shown in Fig. \ref{fig:quantum}(d)--(f). The non-monotonicity begins nearly after $n = 500$ (equivalently $t \simeq 7.96$), $n = 400 $ (equivalently $t \simeq 19.10$), and $n = 300$ (equivalently $t \simeq 23.88$ ) respectively.  In contrast, the von Neumann entropy increases from zero and shows a maximum at $\varepsilon = 0.5$ irrespective of the value of $\tau$, as shown in Fig. \ref{fig:quantum}(g)--(i). Which means that maximum decoherence of individual systems and entanglement between two systems occur at $\varepsilon = 0.5$. But from the Fig. \ref{fig:quantum} (a)--(c) we conclude that the  minima (maxima) of interaction energy (the von Neumann entropy) does not imply $\Delta E = 0$ in general. We take $\tau = 0.3$ as a 
	representative case to illustrate our results in detail.
	\begin{figure}[h]
		\subfigure[]{
			\includegraphics[width=0.22\textwidth]{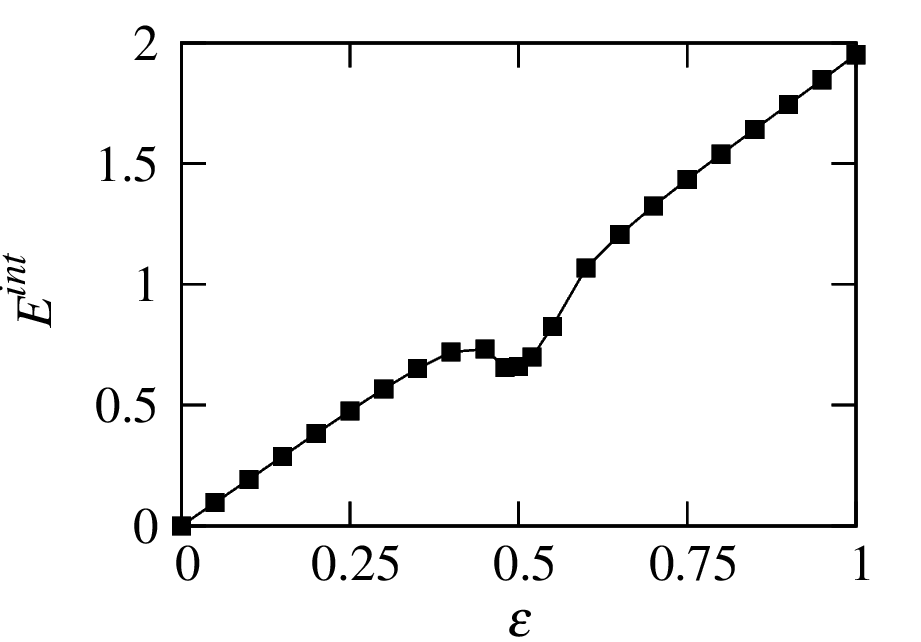}
		}
		\subfigure[]{
			\includegraphics[width=0.22\textwidth]{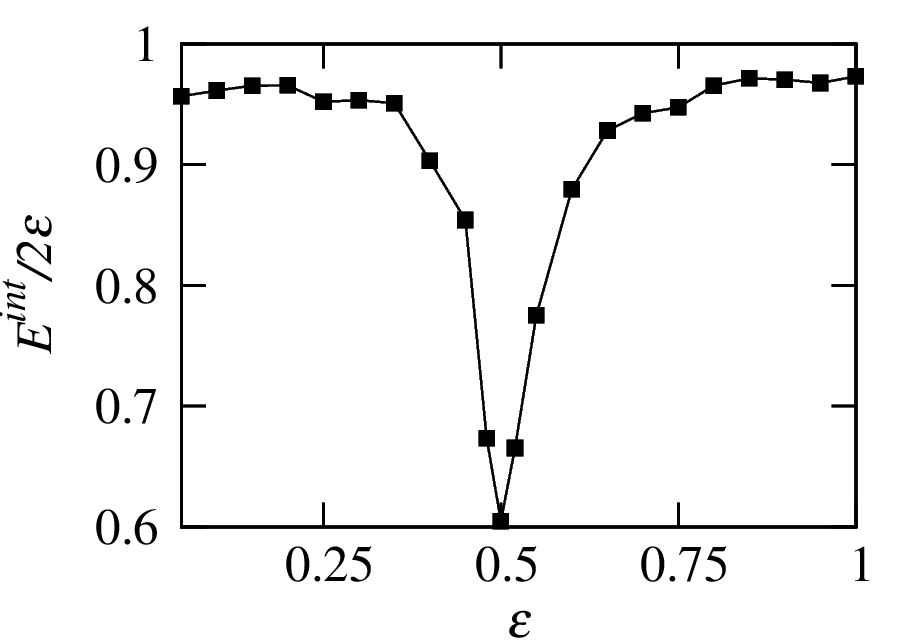}
		}
		\subfigure[]{
			\includegraphics[width=0.22\textwidth]{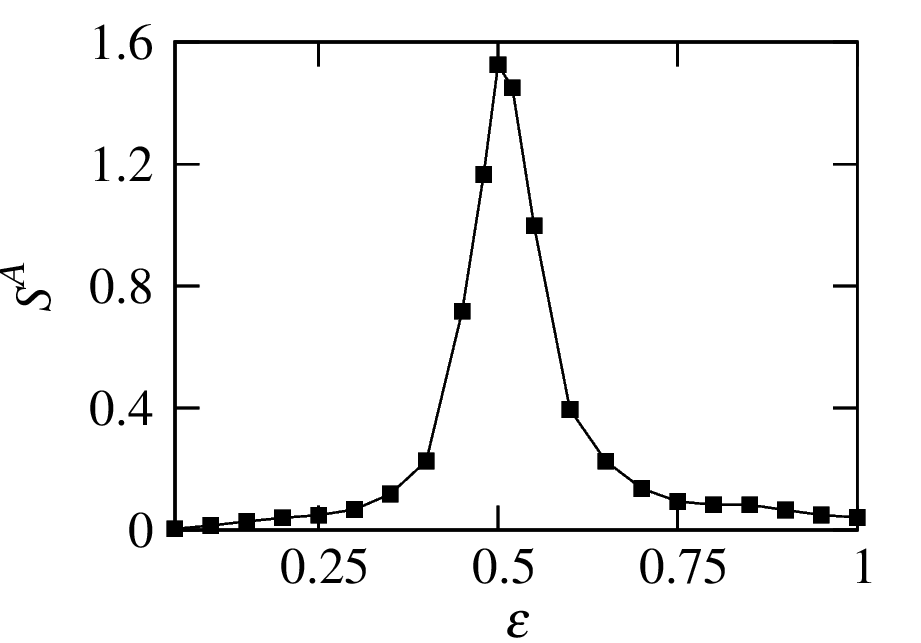}
		}
		\subfigure[]{	
			\includegraphics[width=0.22\textwidth]{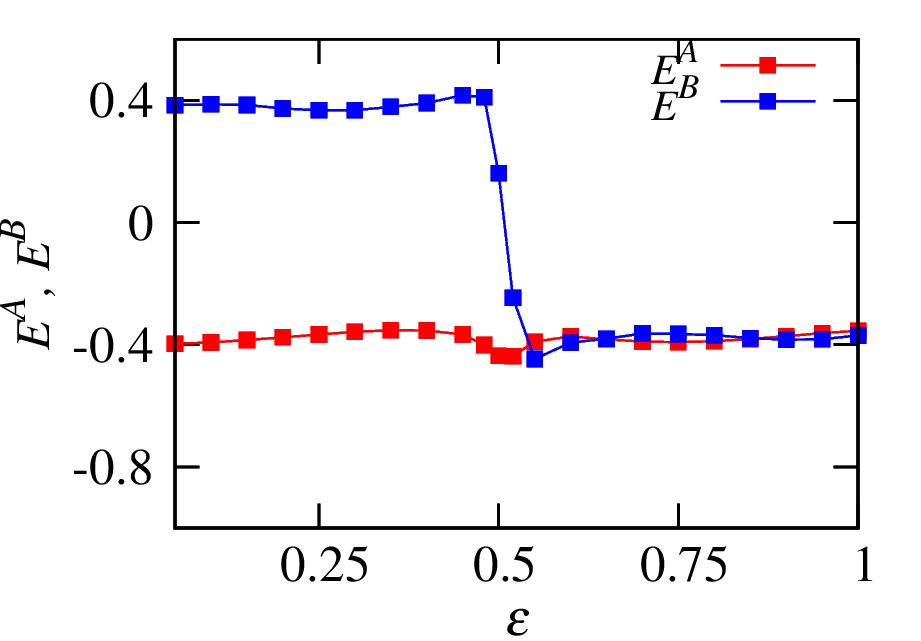}
		}
		\caption{\emph{(color online)} (a) The average interaction energy ($E^{\rm int}$), (b) the average interaction energy divided by twice the coupling constant ($E^{\rm int}/2\varepsilon$), (c) the von Neumann entropy ($S^A$), (d) the average individual energies $E^A$ and $E^B$ as a function of the coupling constant $\varepsilon$ after 1000 kicks for kicking period parameter $\tau = 0.3$.}
		\label{fig:quantum_tau_p3}
	\end{figure}

	We take $\tau = 0.3$ as a representative case to illustrate our results in detail. We have shown the behaviour of the quantities $E^{\rm int}$, $E^{\rm int}/2\varepsilon$, $S^A$, $E^A$, and $E^B$ as a function of $\varepsilon$ after  $1000$ kicks in Fig.~\ref{fig:quantum_tau_p3}.  As shown in Fig.~\ref{fig:quantum_tau_p3} (a), (b), and (c) all the three quantities $E^{\rm int}$, $E^{\rm int}/2\varepsilon$, and $S^A$ show a non analytic behaviour at $\varepsilon = 0.5$---which confirms the transition there. In  Fig.~\ref{fig:quantum_tau_p3}(d) energies of the individual systems suddenly coincide with each other above $\varepsilon = 0.5$. This means the individual systems share their energies equally indicating the quantum synchronization. 
	\subsection{ Inter-system single-qubit mutual information}
	\label{single_cubit_mi}
	\begin{figure*}[htpb!]
		%
		\subfigure[]{%
			\label{fig:first}
			\includegraphics[width=0.30\textwidth]{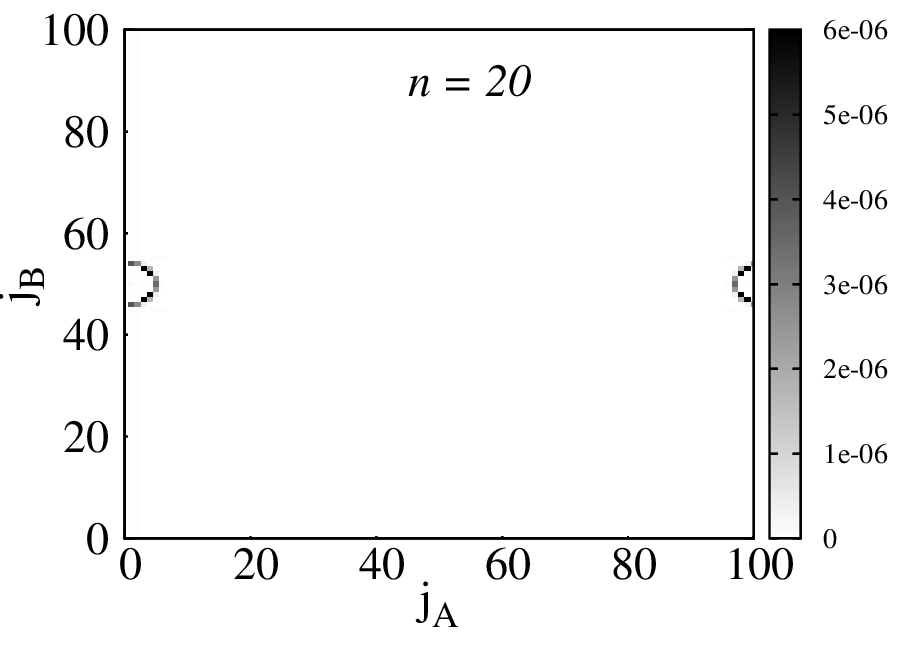}
		}%
		\subfigure[]{%
			\label{fig:second}
			\includegraphics[width=0.30\textwidth]{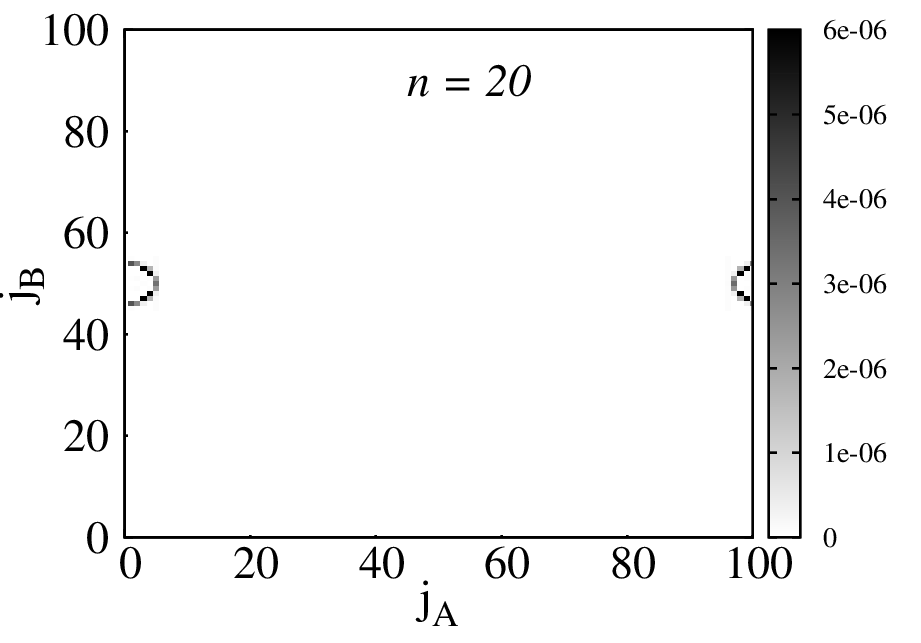}
		}%
		\subfigure[]{%
			\label{fig:first}
			\includegraphics[width=0.30\textwidth]{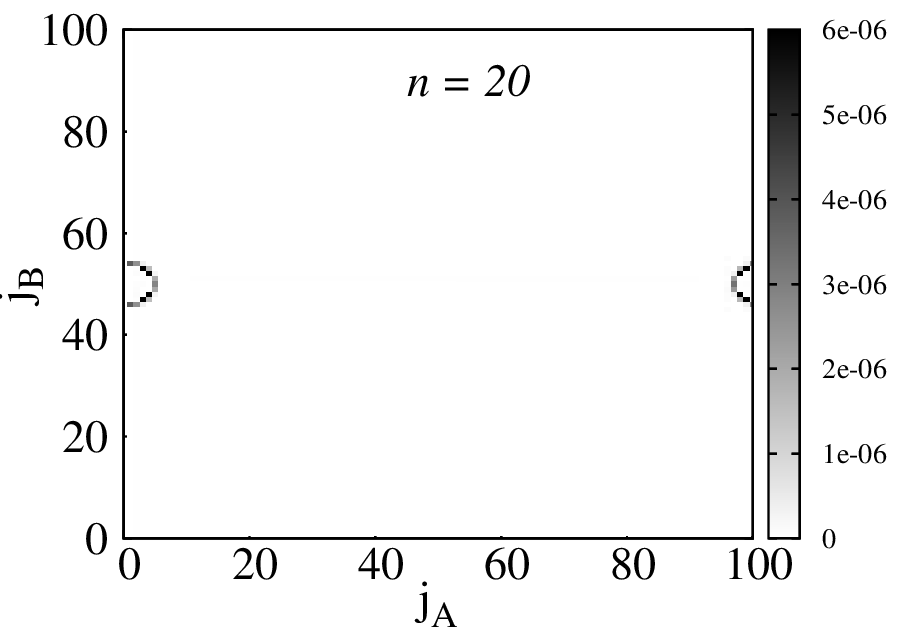}
		}\\%
		\subfigure[]{%
			\label{fig:first}
			\includegraphics[width=0.30\textwidth]{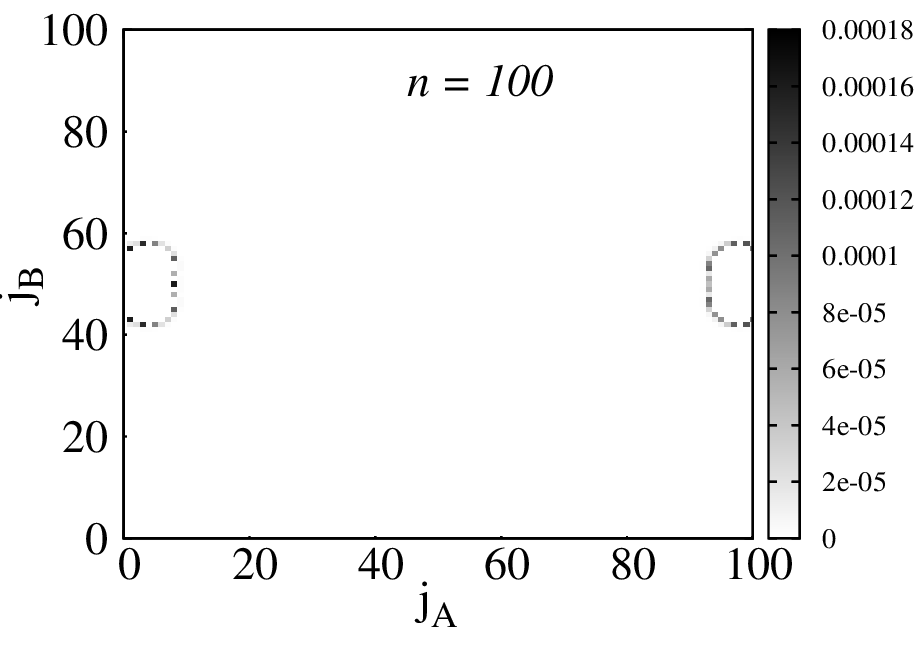}
		}%
		\subfigure[]{%
			\label{fig:second}
			\includegraphics[width=0.30\textwidth]{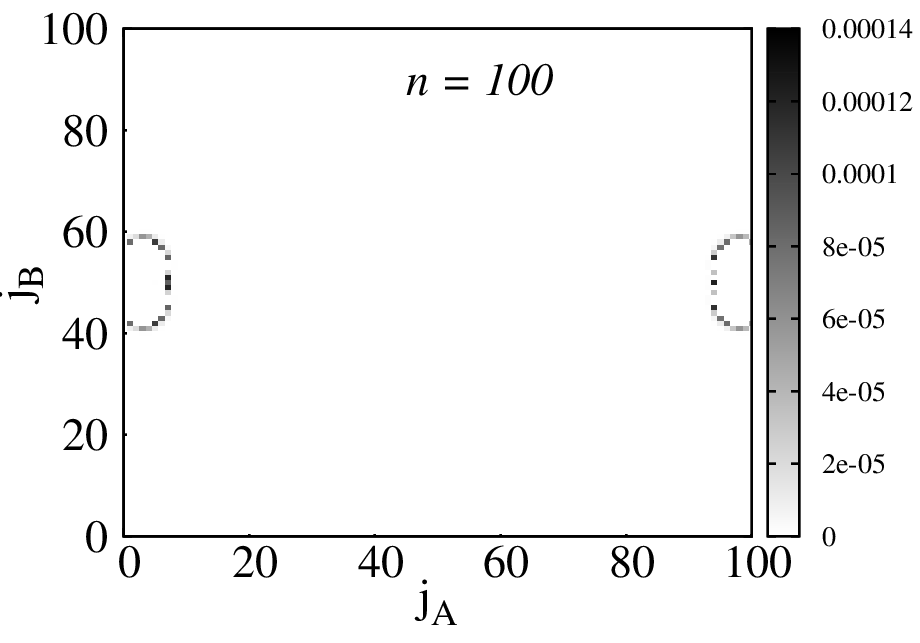}
		}%
		\subfigure[]{%
			\label{fig:first}
			\includegraphics[width=0.30\textwidth]{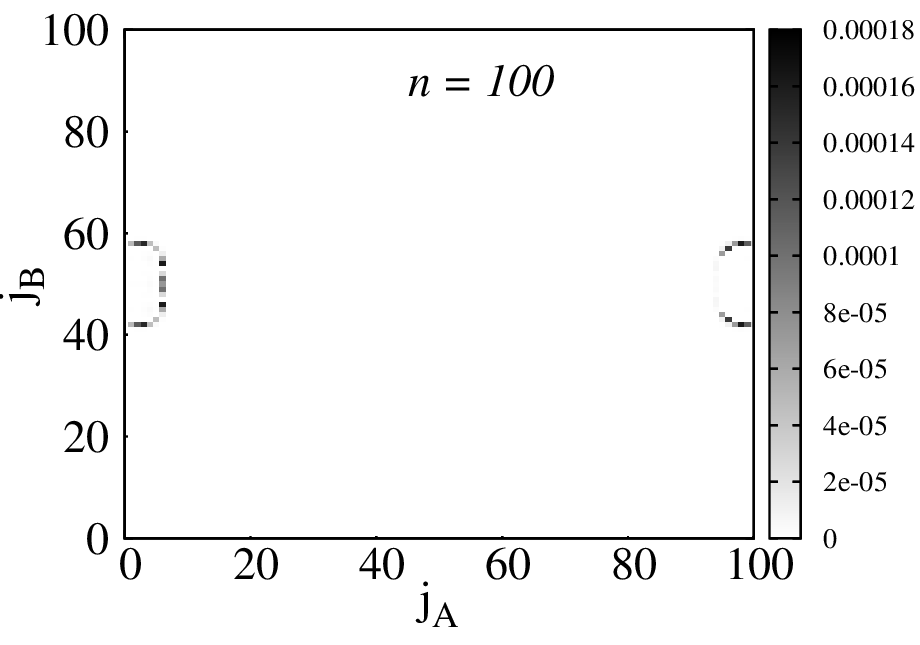}
		}\\%
		\subfigure[]{%
			\label{fig:second}
			\includegraphics[width=0.30\textwidth]{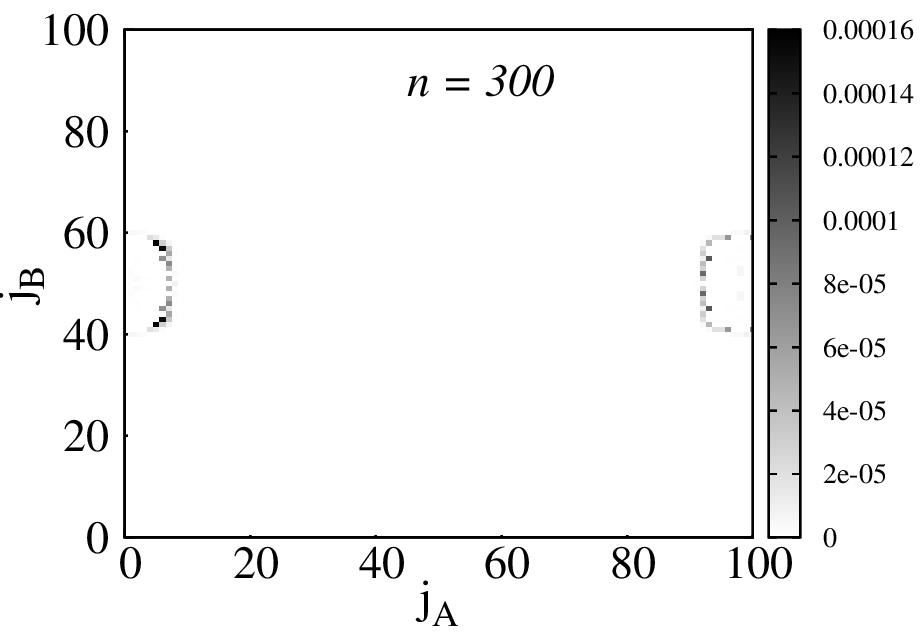}
		}%
		\subfigure[]{%
			\label{fig:second}
			\includegraphics[width=0.30\textwidth]{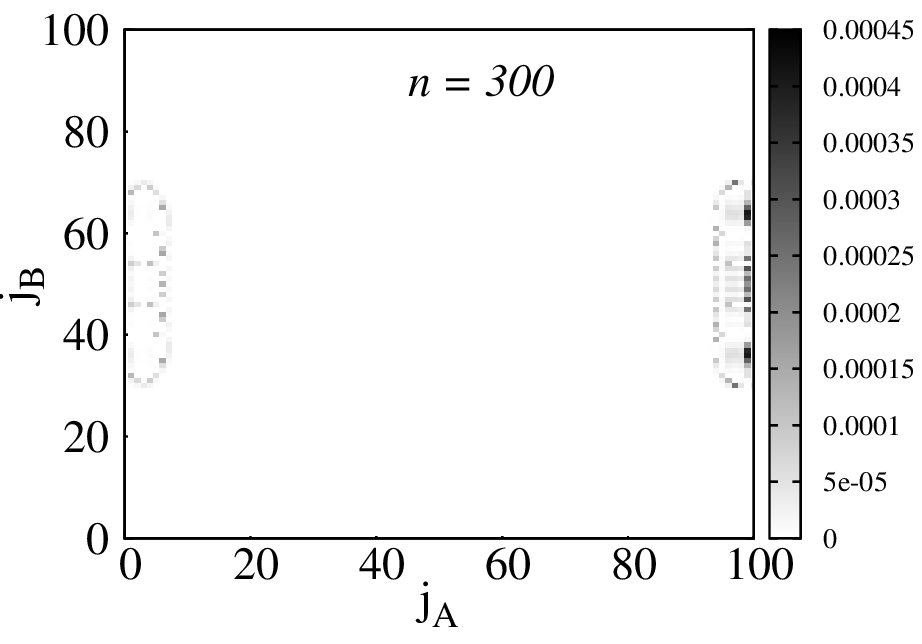}
		}%
		\subfigure[]{%
			\label{fig:second}
			\includegraphics[width=0.30\textwidth]{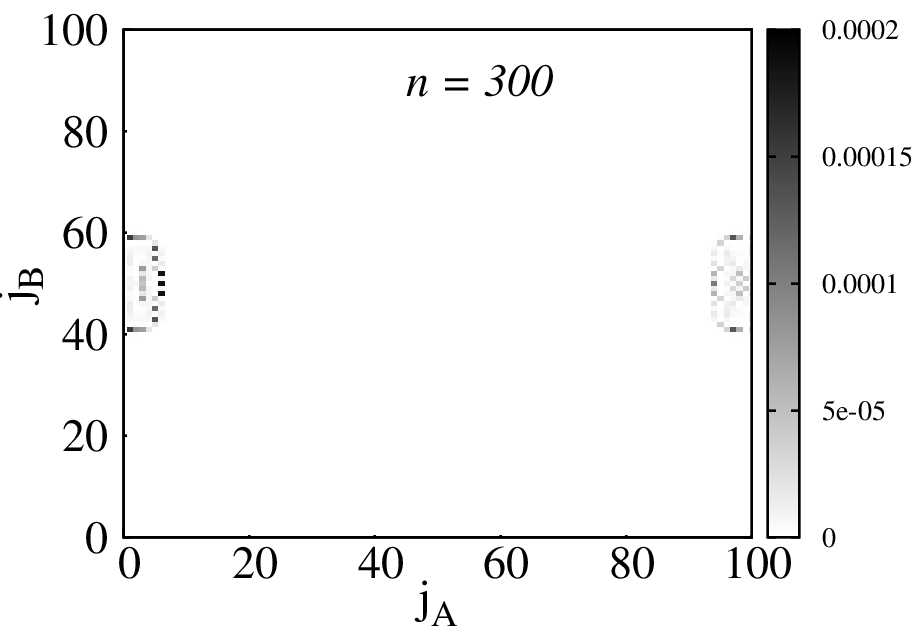}
		}\\%
		\subfigure[]{%
			\label{fig:first}
			\includegraphics[width=0.30\textwidth]{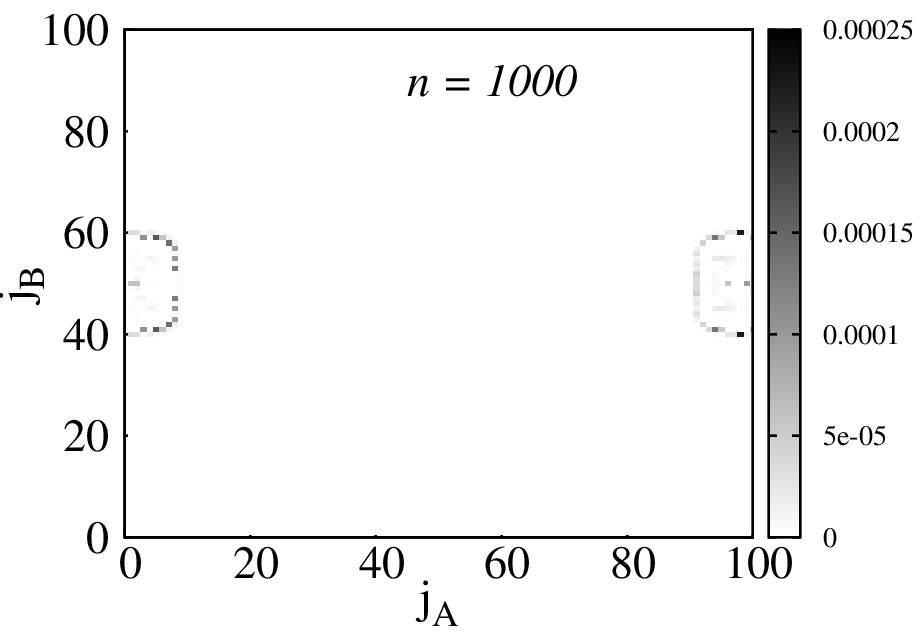}
		}%
		\subfigure[]{%
			\label{fig:second}
			\includegraphics[width=0.30\textwidth]{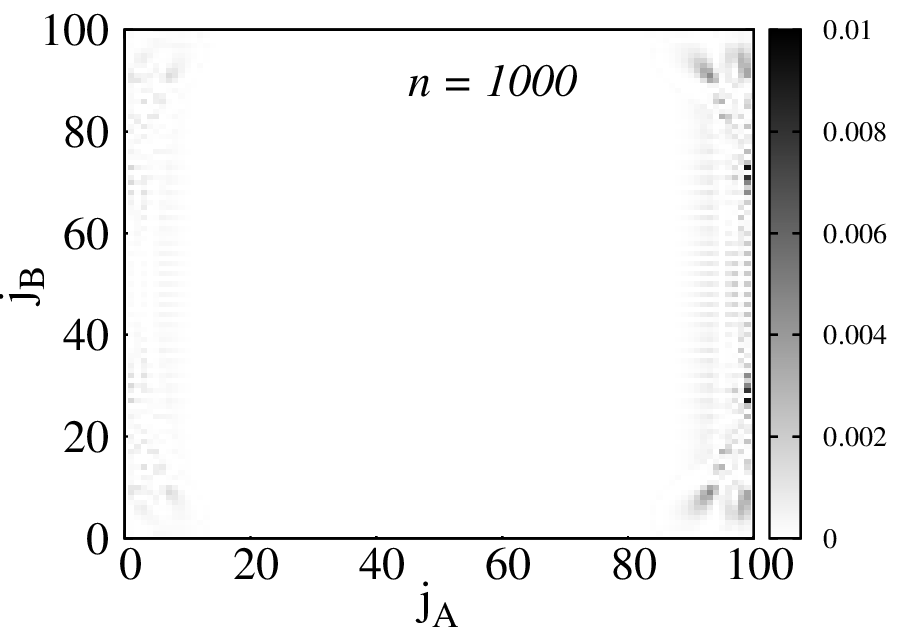}
		}%
		\subfigure[]{%
			\label{fig:first}
			\includegraphics[width=0.30\textwidth]{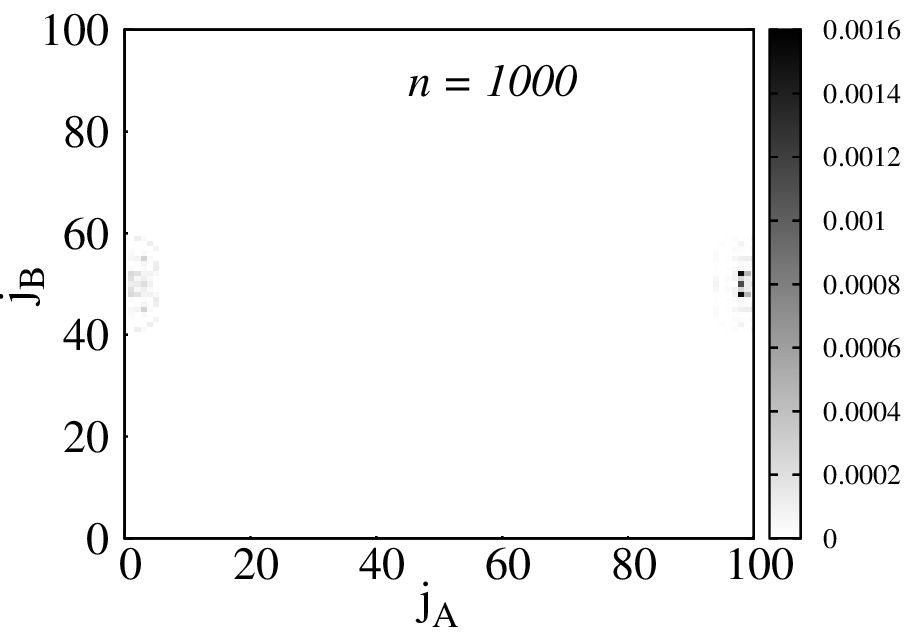}
		}\\%
		\caption{\label{fig:fig_1}{The mutual information between $j^{th}_A$ qubit in system $A$ and $j^{th}_B$ qubit in system $B$ for different coupling strength $(\varepsilon)$ and time $(n)$ :  (a) $\varepsilon = 0.05$, $n = 20$,(b) $\varepsilon = 0.50$, $n = 20$,(c)$\varepsilon = 1.00$, $n = 20$,(d)$\varepsilon = 0.05$, $n = 100$,(e)$\varepsilon = 0.50$, $n = 100$, (f) $\varepsilon = 1.00$, $n = 100$, (g) $\varepsilon = 0.05$, $n = 300$, (h) $\varepsilon = 0.50$, $n = 300$, (i) $\varepsilon = 1.00$, $n = 300$, (j) $\varepsilon = 0.05$, $n = 1000$, (k) $\varepsilon = 0.50$, $n = 1000$, (l) $\varepsilon = 1.00$, $n = 1000$. \textcolor{black}{In all subplots, the abscissa denotes the actual position of the $j^{th}_A$ qubit in system A and the ordinate denotes the actual position of the $j^{th}_B$ qubit in system B.} Initial state: $x_0 = 1$, $y_0 = N/2$. Kicking interval parameter: $\tau = 0.3$.}
		}%
		\label{fig:subfigures}
		\label{fig:quantum_mi} 
	\end{figure*}
	
	\textcolor{black}{It is also interesting to investigate the correlation between individual qubits in system $A$ and system $B$ in order to explain the dynamical phase transition that occurs at $\varepsilon = 0.5$. We take $\tau = 0.3$ as our representative case to show the time dependence of mutual information between $j_A$ th qubit in system $A$ and  $j_B$ th qubit in system $B$ in Fig.~\ref{fig:quantum_mi}. Initially, every qubit in each system in uncorrelated. As time evolves they get correlated and the locus of  maximally correlated qubits spreads out with time  circularly in $(j_A,j_B)$ plane centering the initial state, i.e., $(1,N/2)$. As shown in Fig.~\ref{fig:quantum_mi} (a), (d), (g), (j) and  \label{fig:quantum_mi}(c), (f), (i), (l)  for $\varepsilon = 0.05$ and $\varepsilon = 1.00$ respectively, after a certain time it stops growing due to the convex light cone structure of the Green function \citep{sur18_2}. Even after a long time maximally correlated pairs reside inside a circle centering the 
point 
		$(1,N/2)$ for $\varepsilon = 0.05$ and $\varepsilon = 1.0$ as shown in Fig.~\ref{fig:quantum_mi} (j) and (l) respectively. In contrast, for $\varepsilon = 0.5$ this locus no longer remains circular after a certain time and spreads in the $(j_A,j_B)$ plane as shown in Fig.~\ref{fig:quantum_mi} (h) and (k). Hence, the spreading of quantum correlation is a signature for the transition. Not only the mutual information (twice of $S^A$) between the systems $A$ and $B$ is maximum at $\varepsilon = 0.5$ as seen from Fig.~\ref{fig:quantum_tau_p3}(c) but also maximum value of the mutual information between individual qubits is also much higher. As seen from Fig.~\ref{fig:quantum_mi} (j), (k), and  (l) the maximum value  of mutual information is $0.00025$ for $\varepsilon = 0.05$, is $0.01$ for $\varepsilon = 0.5$, and is $0.0016$ for $\varepsilon = 1.0$ respectively.}
	\subsection{A discussion on Local coupling}
	In this context we want to discuss another type of coupling scheme. The coupling discussed above is global in the sense that every qubit in system $A$ interacts with every qubit in system $B$. Instead the coupling can be local; the limiting case would be the coupling occurs between the two systems $A$ and $B$ only through individual qubits.  In this scheme the $j^{th}$ spin in system $A$ will get coupled only with the $j^{th}$ spin in system $B$ via the same time and space dependent potential. For further mention this will be named as `local coupling'. The coupling Hamiltonian can be obtained from Eq.~\ref{eq:hamil_ab2} by introducing a Kronecker delta function and is given by, 
	\begin{eqnarray}
	{H}^{AB}(t) =  \frac{\varepsilon}{2}\sum^N_{j_A,j_B=1} \delta_{j_A,j_B} \cos(\frac{2 \pi j_A}{N}) \cos(\frac{2 \pi j_B}{N})\sigma^z_{j_A}  \sigma^z_{j_B} \nonumber\\ \sum^{\infty}_{n =-\infty} \delta(\frac{2\pi t }{\tau} -n).\nonumber\\
	\end{eqnarray}
	This type of coupling has no `classical' analog; in the sense that the coupling  Hamiltonian cannot be factorized in two parts corresponding to systems $A$ and $B$ as given in Eq.~\ref{eq:hamil_ab}. The joint state of system $A$ and $B$ is still given by Eq.~\ref{eq:final_st}, but the wave function of the joint state after $n$ kicks will be given by,

		\begin{eqnarray}
		&\tilde{G}^{x_n;y_n}_{x_0;y_0}(n) = \sum_{x_1,x_2,...,x_{n-1}} \sum_{y_1,y_2,...y_{n-1}} \prod^{n-1}_{i=0}G^{x_{i+1}}_{x_i}(\frac{\tau}{2 \pi}) G^{y_{i+1}}_{y_i}(\frac{\tau}{2 \pi})\nonumber\\& \times  e^{i (g+\varepsilon(1-\delta_{x_{i+1},y_{i+1}}))(\cos(\frac{2 \pi {x_{i+1}}}{N})+\cos(\frac{2 \pi y_{i+1}}{N}))}.
		\end{eqnarray} 

	As seen in the above equation the joint Green function behaves like two uncoupled systems with an effective site dependent potential strength parameter $g+\varepsilon(1-\delta_{x_{i+1},y_{i+1}})$. This makes the joint Green function very different from the same given in Eq. \ref{eq:final_st}. Here, the interaction energy after $n$ kicks is given by,

	\begin{eqnarray}
	E^{\rm int}_{n}  = \varepsilon \frac{N}{4} - \varepsilon \sum_{x_n,y_n} |\tilde{G}^{x_n;y_n}_{x_0;y_0}(n)|^2 (\cos(\frac{2 \pi x_n}{N})+\cos(\frac{2 \pi y_n}{N})) \nonumber\\
	(1-\delta_{x_n,y_n}). \nonumber\\
	\end{eqnarray}

\begin{figure*}[htpb!]
		\begin{center}
			\subfigure[]{%
				\label{fig:second}
				\includegraphics[width=0.26\textwidth]{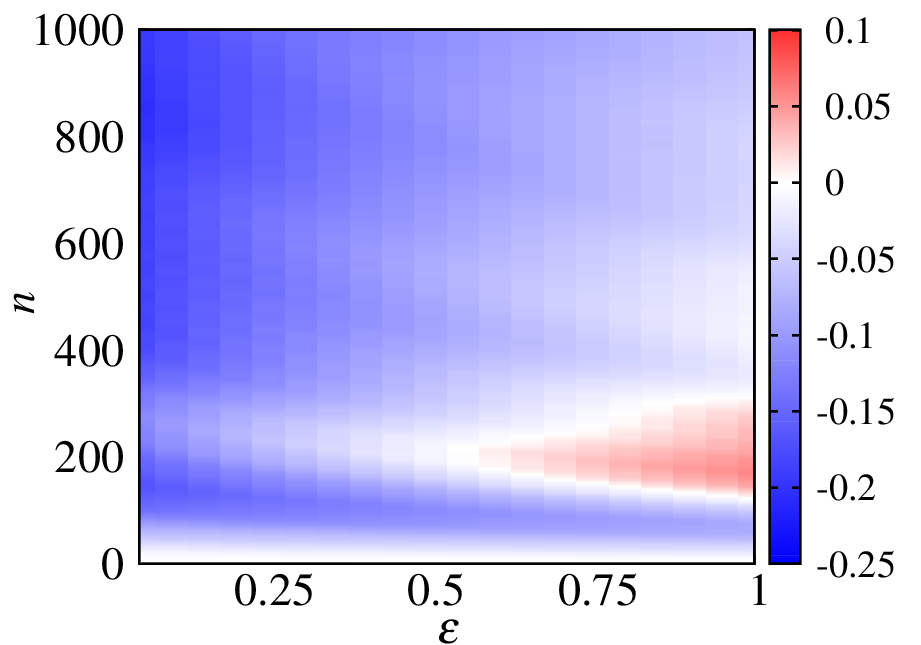}
			}%
			\subfigure[]{%
				\label{fig:second}
				\includegraphics[width=0.26\textwidth]{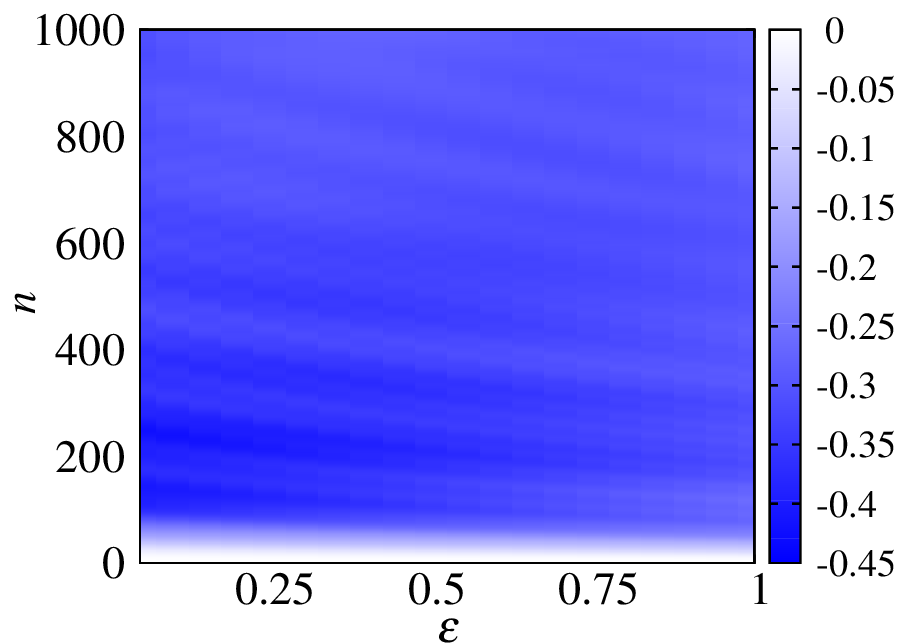}
			}%
			\subfigure[]{%
				\label{fig:second}
				\includegraphics[width=0.26\textwidth]{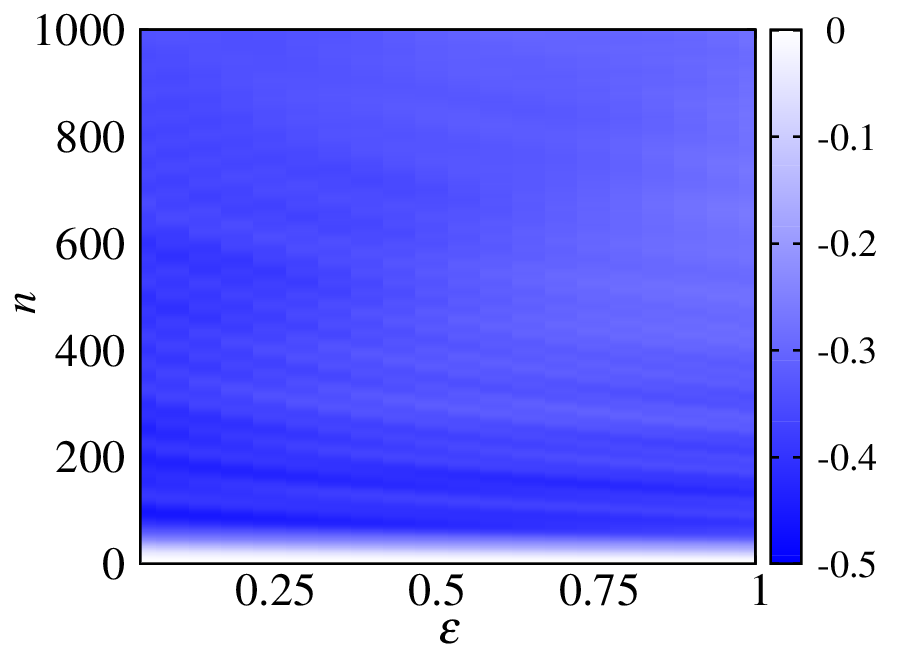}
			}\\%
			\subfigure[]{%
				\label{fig:second}
				\includegraphics[width=0.26\textwidth]{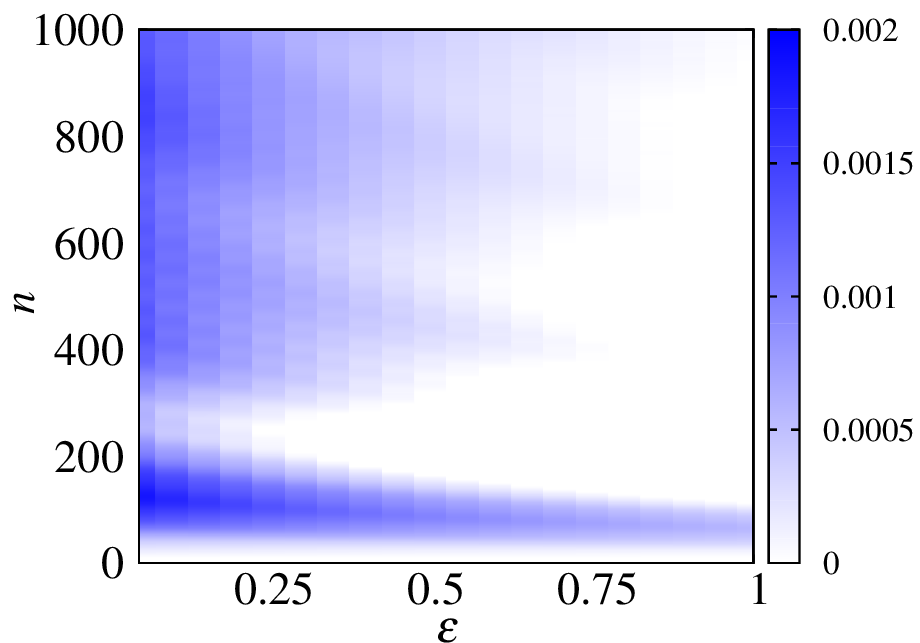}
			}%
			\subfigure[]{%
				\label{fig:second}
				\includegraphics[width=0.26\textwidth]{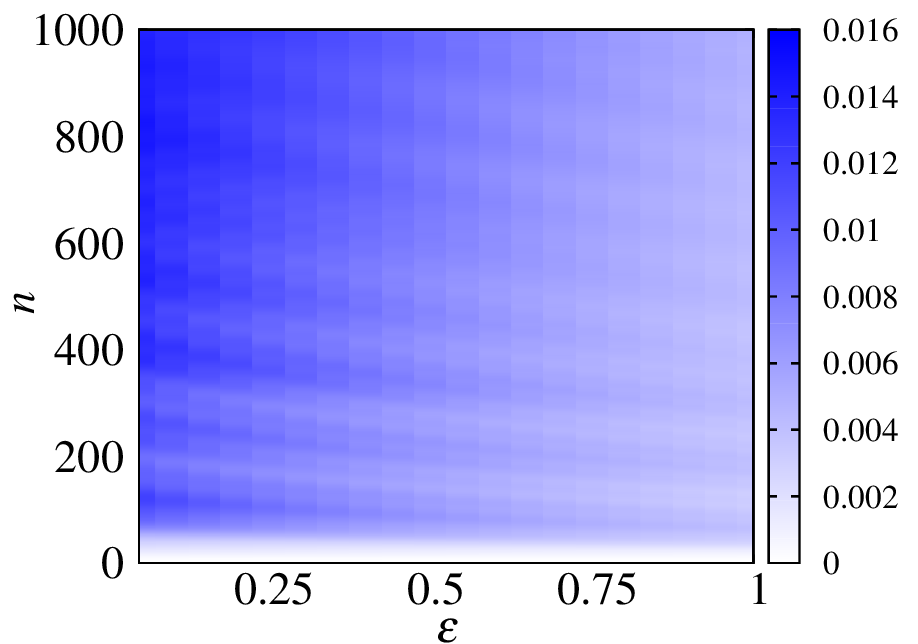}
			}%
			\subfigure[]{%
				\label{fig:second}
				\includegraphics[width=0.26\textwidth]{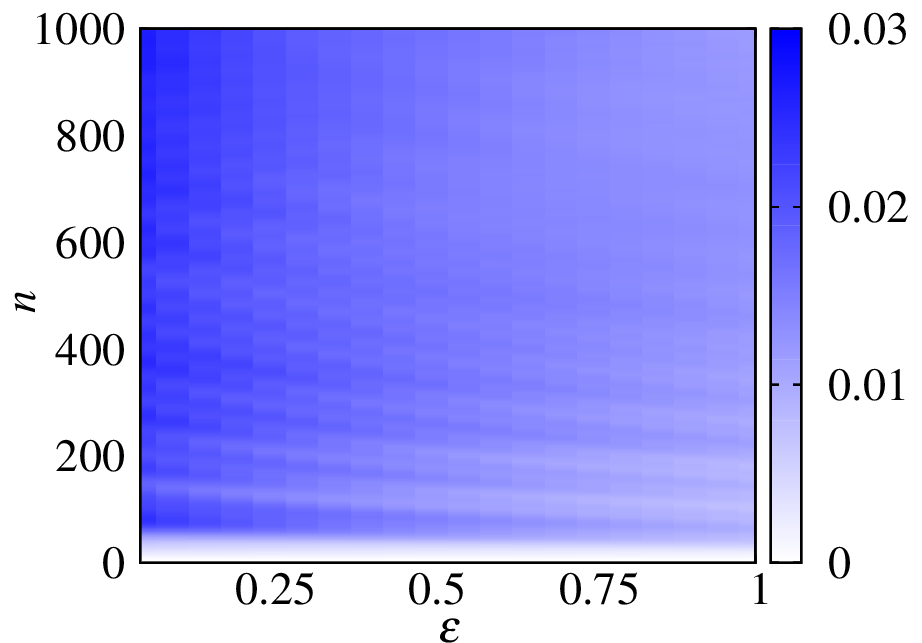}
			}\\%
			\subfigure[]{%
				\label{fig:second}
				\includegraphics[width=0.26\textwidth]{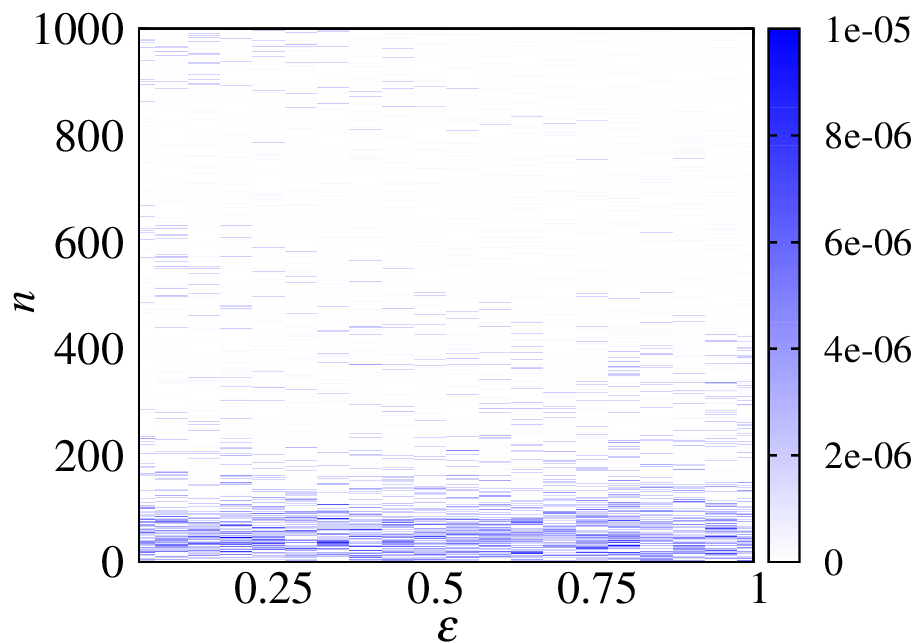}
			}%
			\subfigure[]{%
				\label{fig:second}
				\includegraphics[width=0.26\textwidth]{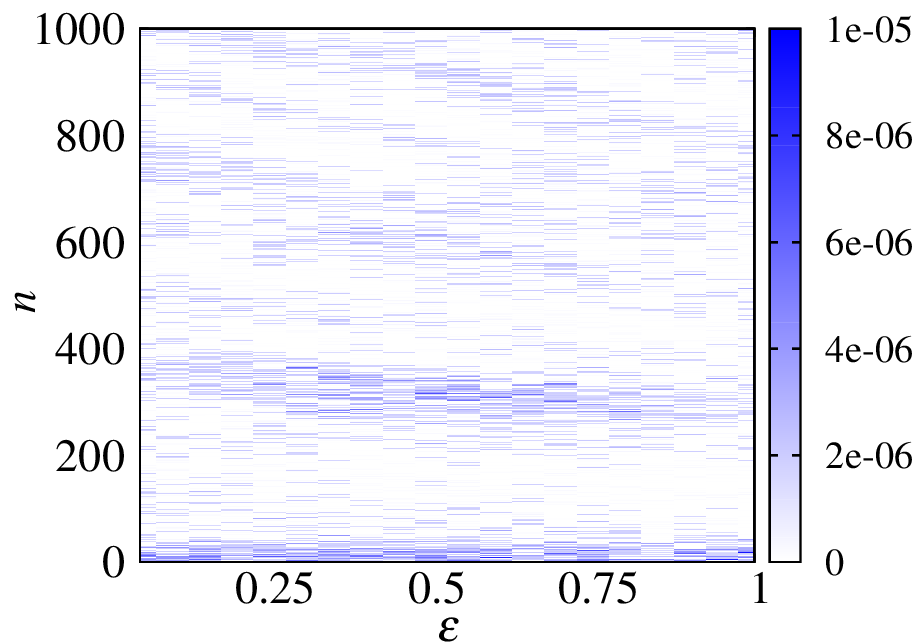}
			}%
			\subfigure[]{%
				\label{fig:second}
				\includegraphics[width=0.26\textwidth]{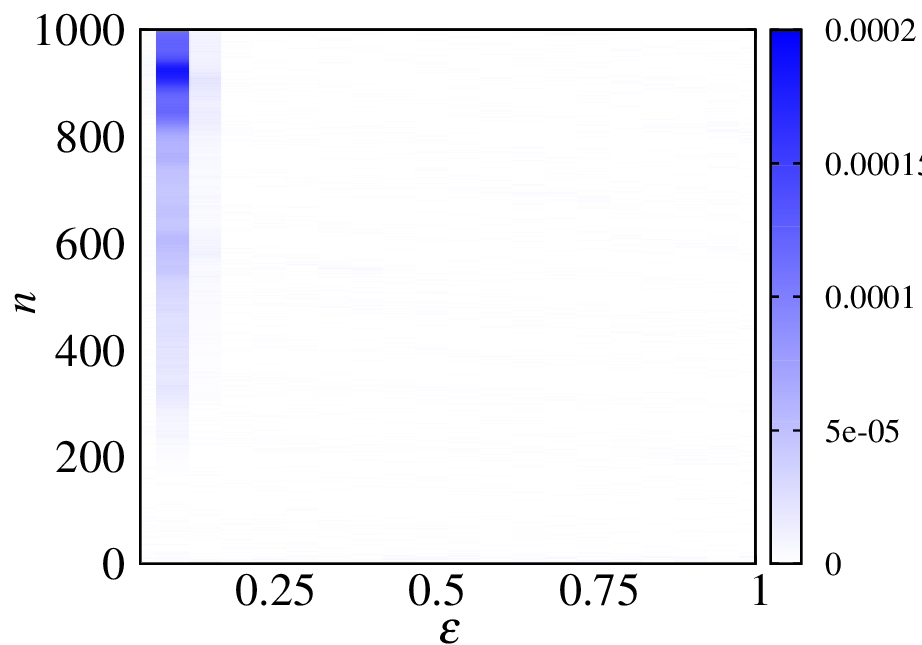}
			}\\%
		\end{center}
		\caption{\label{fig:fig_1}{\emph{(color online)} Local coupling scheme: The difference between the energies of system $A$ and $B$ ($\Delta E = E^A - E^B$) as a function of number of kicks ($n$) and the coupling parameter ($\varepsilon$) for kicking intervals (a)$0.1/2\pi$ (b)$0.3/2\pi$  (c)$0.5/2\pi$. The average interaction energy divided by the coupling parameter ($E^{\rm int}/\varepsilon$) as a function of number of kicks ($n$) and the coupling parameter ($\varepsilon$) for kicking intervals (d)$0.1/2\pi$ (e)$0.3/2\pi$  (f)$0.5/2\pi$. The von Neumann entropy ($S^A$ or $S^B$) as a function of number of kicks ($n$) and the coupling parameter ($\varepsilon$) for kicking intervals (g)$ 0.1/2\pi$ (h)$0.3/2\pi$  (i)$0.5/2\pi$. Number of qubits in each system: $N = 100$. Initial state: $x_0 = 1$, $y_0 = N/5$.  }
		}%
		\label{fig:subfigures}
		\label{fig:quantum_nn}   
	\end{figure*}

	Fig.~\ref{fig:quantum_nn} show the time dependence of  difference between average energies of individual systems ($\Delta E = E^A -E^B$), average interaction energy divided by the coupling parameter ($E^{\rm int}/\varepsilon$) between the systems $A$ and $B$ and von Neumann entropy ($S^A$) of the systems $A$ or $B$  for different values of $\varepsilon$ and $n$ for the `local coupling' scheme. Here, the initial state plays a significant role in the dynamics because the interaction is local. If the down spin in system $A$ is very far from the same in system $B$ the systems will have effectively no coupling. Hence, unlike the earlier case we choose the initial state $|1,N/5\rangle$ to show our results. However, we have tried with other initial states but the dynamics is not qualitatively very different.
	As shown in Fig.~\ref{fig:quantum_nn}(a)--(c), the difference between the average energies of the individuals systems ($\Delta E$) initially decreases with time independent of the value of $\varepsilon$ and $\tau$, and mostly remain so. Unlike the earlier coupling scheme here no such branching of $\Delta E$ depending on the coupling is observed. Same kind of trend is seen in in case of $E^{\rm int}/\varepsilon$, as shown in Fig.~\ref{fig:quantum_nn}(d)--(f). Fig.~\ref{fig:quantum_nn}(g)--(i) show the von Neumann entropy ($S^A$) is very small compared to the earlier case. This means the coupling is negligible or in other words, the do not get entangled appreciably. Only thing we can say that $S^A$ increases with kicking period.
	\section{Conclusions}
	In conclusion, we have studied the synchronization for coupled Harper systems. We extended the concept of synchronization from classical to quantum scenario for the same coupled systems. To do so, we have investigated through the method of average interaction energy between the participating subsystems in both classical and quantum cases. Further, we have followed different paths also to study the synchronization. 
	In this paper, to make the analogy between the classical and the quantum scenarios, we have illustrated our results explicitly for $\tau = 0.3$ for the classical part. In Fig.~\ref{fig:classical}, \ref{fig:classical_jpd}, \ref{fig:classical_int_en}, \ref{fig:classical_tau_p1}, and \ref{fig:classical_tau_p5} all plots are done using a single set of initial conditions $(x^{A}(0), p^{A}(0), x^{B}(0), p^{B}(0)) = (0.5, 0.4, 0.3, 0.5)$. However, one may obtain unaltered conclusions with other set of initial conditions. Here, we adopt the joint probability density technique, in addition to the average interaction energy method, to detect the synchronized (desynchronized) state. But, choosing of different $\tau$ may not lead to observe the transition between the synchronized and the desynchronized states, though there is a kink observed in the average interaction energy. For example, at $\tau = 0.1$, the discontinuity is observed at $\varepsilon = 0.5$ (see Fig.~\ref{fig:classical_tau_p1}(a) or (b)), but the 
difference in the 
	joint probability density does not confirm this transition (see Fig.~\ref{fig:classical_tau_p1}(c) and (d)). Similar kind of observations one can get for any $\tau \lesssim 0.3$. Further, for $\tau \gtrsim 0.5$, since the intrinsic subsystems show the global chaotic nature and they are always in synchronized state independent of the coupling strength. This leads to make us the possible conclusion that \emph{the transition is always associated with the kink, but the converse is not always true}. Further, one possible future direction may be the explicit study of the dynamics of the coupled subsystems in the desynchronized state, which seems to us the existence of local chaos~\citep{walker69}. However, in the synchronized state, the participating subsystems are full fledged chaotic in nature. 
	In quantum scenario, we observe a dynamical phase transition at $\varepsilon = 0.5$ irrespective of the kicking period. Where, $\varepsilon = 0.5$ is the transition point from desynchronized state to MS state in classical context for $\tau \gtrsim 0.3$. In this regime two quantum systems $A$ and $B$ equal their energies by sharing though the coupling for $\varepsilon > 0.5$. For the case $\tau \lesssim 0.3$ the quantum systems instead show a energy level crossing at $\varepsilon = 0.5$. We do not see any transition in the classical context also as mentioned already. The average interaction energy between the systems $A$ and $B$ divided by the coupling constant $\varepsilon$ shows a minimum at $\varepsilon = 0.5$ irrespective of the kicking period. This is a common feature which is seen in both classical and quantum scenario. Moreover, the quantum correlation measures, viz., the von Neumann entropy shows a maximum at $\varepsilon = 0.5$. So, the transition is associated with a much larger value of 
	entanglement between the systems $A$ and $B$, or in other words, more decoherence in the individual systems. So, the transition from desynchronized to synchronized state can be thought of as a classical manifestation of a dynamical phase transition in quantum many body system. A small discussion on inter system single qubit correlation is given in Section~\ref{single_cubit_mi}. It can be seen that quantum correlations spread over the systems at the transition point.
	We studied the quantum scenario and discussed the results in joint Green function formalism, i.e., starting from one down spin(or, one magnon) in each system. Since, system is non interacting,  the time-dependent wave function from any initial state can be written as a product of the joint Green function. Although a state with more number of down spins in each system will result in much stronger coupling the qualitative features of our results  will not change.
	We have not commented about the nature of the transition in both classical and quantum scenarios. Though we have analytical expressions for the time dependent wave functions, energies, etc., in quantum dynamics---we need to check the analyticities of the energies, entropies or other correlations at the transition point. Hence determining the nature of the transition is a challenging analytical problem.
	Finally, the local coupling scheme which we have discussed does not give rise to any such transitions. The possible reasons might be the coupling has no classical analog, i.e., the coupling Hamiltonian cannot be written as a product of two terms corresponding to the systems $A$ and $B$ and the coupling is very weak. It seems that the transition is a `classical' phenomena, which cannot be obtained through local coupling. This requires further investigation.
	\section*{Acknowledgements}
	The authors thank S. Chakraborty, A. Lakshminarayan, V. Subrahmanyam, and H. Wanare for fruitful discussions. S.S. acknowledges the financial
	support from CSIR, India.
\section*{References}
\bibliographystyle{elsarticle-harv} 
\bibliography{Sur_Ghosh_Manuscript.bib}
\end{document}